\documentclass[floatfix,aps,twocolumn,prx,superscriptaddress,10pt]{revtex4-2}
\usepackage{paperpreamble} 

\definecolor{Nathanblue}{rgb}{0.,0.24,0.51}
\newcommand{\blue}[1]{{\color{Nathanblue}#1}}

\begin{document}

\author{Amit Vashisht}
\thanks{These three authors contributed equally}
\author{Ivan Amelio}
\thanks{These three authors contributed equally}
\author{Laurens Vanderstraeten}
\thanks{These three authors contributed equally}
\affiliation{CENOLI, Universit\'e Libre de Bruxelles, CP 231, Campus Plaine, B-1050 Brussels, Belgium}
\author{Georg M.\ Bruun}
\affiliation{Department of Physics and Astronomy, Aarhus University, Ny Munkegade, DK-8000 Aarhus C, Denmark}
\author{Oriana \surname{K. Diessel}}
\affiliation{ITAMP, Center for Astrophysics, Harvard \& Smithsonian, Cambridge, Massachusetts 02138, USA}
\affiliation{Department of Physics, Harvard University, Cambridge, Massachusetts 02138, USA}
\author{Nathan Goldman}
\email{nathan.goldman@lkb.ens.fr}
\affiliation{CENOLI,
Universit\'e Libre de Bruxelles, CP 231, Campus Plaine, B-1050 Brussels, Belgium}
\affiliation{Laboratoire Kastler Brossel, Coll\`ege de France, CNRS, ENS-PSL University, Sorbonne Universit\'e, 11 Place Marcelin Berthelot, 75005 Paris, France
}

\title{\blue{Chiral polaron formation on the edge of topological quantum matter}}

\begin{abstract}

Immersing a mobile impurity in a quantum many-body environment can reveal fundamental properties of the background medium, hence providing a powerful probe of quantum matter. This approach is particularly intriguing when considering media with exotic properties, such as strongly-correlated phases and topological states of matter. In this work, we study the dressing of a mobile impurity interacting with a chiral mode, as provided by the edge of topological quantum matter. The resulting ``chiral polaron'' is characterized by an asymmetric spectral function, which reflects the chirality and group velocity of the topological edge mode and the drag experienced by the mobile impurity. We first build our theoretical understanding from an effective one-dimensional chiral model, which captures the hallmark signatures of the chiral polaron. We then demonstrate how this simple picture extends to realistic models of integer and fractional Chern insulator states, by adapting tensor-network methods to polaron spectroscopy. Injecting mobile impurities on the edge of topological quantum matter is shown to be a powerful tool to probe exotic edge properties, particularly suitable for cold-atom experiments.

\end{abstract}

\date{\today}

\maketitle

\section{Introduction}

Polarons emerge from the interaction of a mobile impurity with the surrounding many-body medium. Originally introduced to describe the motion of electrons in polar crystals~\cite{landau1933,landaupekar1948, lee1953, frohlich1954, feynman1955}, polarons are key to understanding the properties of materials, including semiconductors \cite{Lindemann_1983}, high-temperature superconductors \cite{Mott_1993}, alkali halide insulators \cite{Popp_1972}, and transition metal oxides \cite{Moser_2013}.

Experimental studies performed using ultracold atomic gases have demonstrated the formation of Fermi and Bose polarons, by tuning the impurity-bath interactions via Feshbach resonances \cite{Schirotzek_2009, Nascimbene_2009, Kohstall_2012, Koschorreck_2012, hu2016, jorgensen2016, Scazza2016, Oppong2019, yan2020,Ness2020,fritsche2021stability}. Besides, Fermi \cite{Sidler_2016} and Bose \cite{Tan_2023} polarons were also realized recently in transition metal dichalchogenide (TMD) heterostructures, where their formation sheds light on the optical excitation spectrum of doped monolayer semiconductors and allows for the study of polaron-polaron interactions \cite{camacho-guardian2018, muir2022, Tan_2023,baroni2024}.

Conventional Fermi and Bose polaron settings involve a mobile impurity interacting with a cloud of non-interacting or weakly-interacting fermions or bosons. Despite the simplicity of the medium, the impurity-bath interactions can lead to rich and correlated many-body states. In particular, whereas most properties of the Fermi polaron are fairly well understood even for strong interactions~\cite{massignan2014}, there are many open questions regarding the role of many-body correlations and bound states for the Bose polaron~\cite{Levinsen2015b,Ardila2019,Levinsen2021,Massignan2021a,Mostaan_2023,Christianen_2024}.

Beyond conventional settings, it is appealing to explore polaron formation in exotic media, such as lattice systems with non-trivial Bloch band properties (topological properties or quantum geometry), or strongly-correlated phases of matter. At a fundamental level, the behavior of a mobile impurity interacting with a non-trivial medium is conceptually interesting, and can potentially lead to rich impurity-dressing mechanisms and unconventional properties. Besides, studying the formation of a polaron can provide useful information on the background medium, such that the polaron can be used as a quantum probe for exotic states of matter. From a theoretical point of view, the investigation of polarons in unconventional and strongly-correlated media is at the dawn, in part due to the complications of treating the bath. So far, pioneering works have dealt with 
fractional Chern insulators~\cite{Grusdt2016,munoz2020anyonic}, 
Fermi superfluids along the BEC-BCS crossover~\cite{Nishida2015,yi2015polarons,Pierce2019,Alhyder2022,amelio2023two-dimensional}, 
excitonic insulators~\cite{amelio2023polaron},  twisted homobilayers~\cite{mazza2022strongly},
the Mott transition~\cite{amelio2024insulator},
interacting bosons on a lattice~\cite{colussi2023lattice,Ding2023,santiagogarcia2024lattice},
kinetic magnetism~\cite{khalaf2022, morera2023high-temperature},
many-body topological states in dimerized chains~\cite{grusdt2019},
Holstein polarons in Luttinger liquids~\cite{Kang2021}, 
Fermi baths with Dirac~\cite{Sorout_2020} and flat band dispersion~\cite{pimenov2024}. 
We also note a few related works on the dressing of optical excitations in Mott insulators~\cite{huang2023mott,huang2023spin} and fractional quantum Hall systems~\cite{grass2020optical,raciunas2018, wang2022, wang2024}. 
Theoretical methods are currently being developed, with variational approaches complementing numerical tools such as Quantum Monte Carlo~\cite{Prokofev_2008,ardila2015impurity,bombin2019two-dimensional,ardila2020strong,sanchezbaena2023universal}, exact diagonalization~\cite{amelio2024edpolaron} and tensor network~\cite{Leskinen2010,kantian2014,massel2013dynamics} techniques. Very recently, experiments in TMD heterostructures \cite{smolenski2021signatures, Zhou2021bilayer, shimazaki2021optical, ciorciaro2023, tao2024} and cold-atom simulators \cite{Koepsell2019,Ji2021,lebrat2024, prichard2024} have started to employ polarons in the study of highly correlated states of matter. 

In this context, dropping a mobile impurity into topological quantum matter has recently emerged as an exciting framework for unconventional polaron physics. Several studies have investigated the binding of a mobile impurity to a quasi-hole excitation in fractional quantum Hall (FQH) states~\cite{Grusdt2016, munoz2020anyonic,grass2020, baldelli2021}:
this setting was shown to be ideally suited to probe the underlying many-body Chern number (a topological invariant associated with the many-body wave function), but also to characterize and manipulate anyons. Besides, the Hall drift of a Fermi polaron immersed in the bulk of a Chern insulator was explored in Refs.~\cite{camacho-guardian2019, pimenov2021}.
Importantly, these previous works concerned the behavior of impurities injected in the \emph{bulk} of topological matter, where the dressing entirely relies on bulk excitations and bulk properties (e.g.~the Berry curvature); furthermore, topology manifests in \emph{transport or scattering} scenarios, requiring fine control over individual impurities and bath excitations (e.g.~the anyons).

\subsection*{Chiral polaron: scope of the paper}

Here, we shift gears and explore the formation of polarons on the \emph{edge} of topological insulating states of matter, setting the focus on their associated \emph{spectroscopic} hallmarks. In this framework, the impurity is mainly dressed by the low-energy chiral edge modes circulating around the sample, resulting in a ``chiral polaron'', as we further explain below.

\begin{figure}[!tb]
    \centering
    \includegraphics[width=\columnwidth]{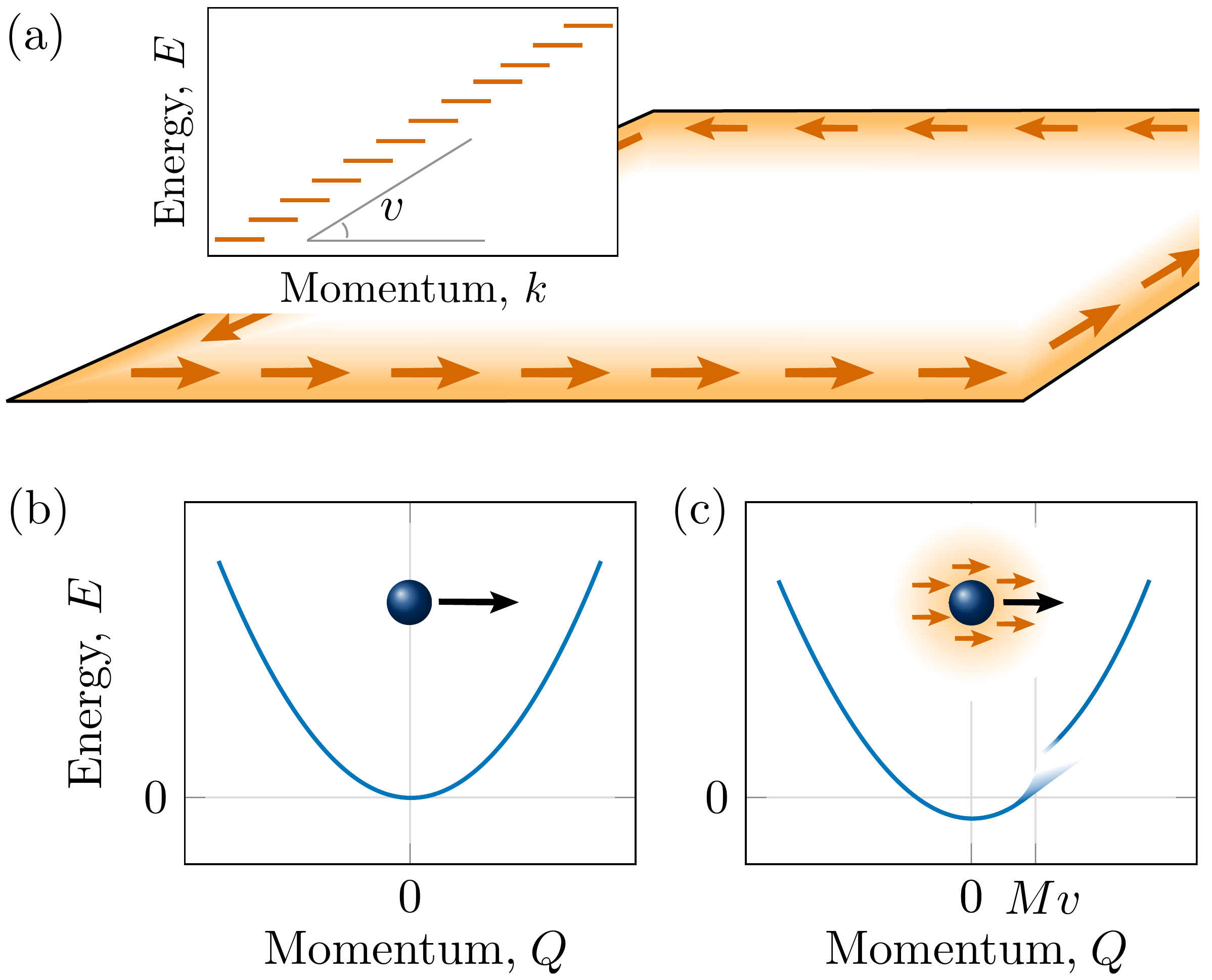}
    \caption{
    {{Chiral dressing of a mobile impurity injected on the edge of a topological insulator}.}
    (a) The edge of the topological insulator provides a chiral mode, with linear dispersion relation $E \sim vk$. (b) We inject a mobile impurity of mass M and momentum $Q$ on the edge of the system. In the absence of interactions with the chiral bath, its dispersion relation is quadratic (free particle). (c) Upon activating interactions, the impurity is dressed by the chiral bath. Its spectral function displays a chiral behavior with strong bath-impurity coupling features at momenta of the order of $Mv$, a clear signature of the chiral dressing.}
    \label{fig:schematic}
\end{figure}

For the sake of concreteness, we consider two-dimensional (2D) Chern insulators, a class of topological insulators characterized by chiral edge modes. This includes systems of non-interacting fermions, filling Bloch bands with non-zero Chern numbers~\cite{thouless1982,haldane1988}, but also the so-called \emph{fractional} Chern insulators, which are strongly-correlated states akin to the FQH states on a lattice~\cite{liu2024}. In the simplest setting, such a topological system displays a single branch of low-energy edge states, with (quasi-)linear dispersion relation $E \sim vk$; see the sketch in Fig.~\ref{fig:schematic}(a). We then inject a mobile impurity of mass $M$ on the edge of this sample. Importantly, we assume that this impurity behaves as a free particle in the absence of coupling to the topological medium, as illustrated in Fig.~\ref{fig:schematic}(b). A chiral polaron then forms upon activating impurity-bath interactions, as signaled by a characteristic behavior of the momentum-resolved impurity spectral function, shown pictorially in Fig.~\ref{fig:schematic}(c). As detailed below, this response function displays two key features:~an asymmetric spectrum with respect to momentum $Q\leftrightarrow -Q$, as a direct consequence of chirality, and a splitting into two broadened lines for momenta larger than the critical momentum $Q_c\!=\!Mv$, reflecting the resonant exchange of energy-momentum between the impurity and the chiral bath.

This phenomenon is explored in this work, through realistic lattice models of Chern insulators and fractional Chern insulators. We also introduce and analyze an effective low-energy model, which correctly captures the key spectral features of the chiral polaron, determined by the dressing of the impurity by low-energy edge excitations. Three theoretical approaches are used to analyse this: exact diagonalisation for small systems,  variational wave function methods, and  tensor network techniques, which are particularly well suited to study the case of fractional Chern insulators, i.e.~a strongly-correlated topological medium. The application of advanced tensor network methods to the polaron context constitutes another key aspect of this work.  

Our results show that the chiral polaron can be used as a quantum probe of topological matter. Indeed, the spectral function that would result from injecting an impurity in the bulk of the sample would not display these key chiral features. Similarly, those signatures would not be present when setting the surrounding medium in a different phase (e.g.~a trivial insulator or a metallic phase). Importantly, the phenomenon is universal, in the sense that it does not rely on the microscopic details of the underlying system. This is demonstrated in this work, by analyzing two topological media of very different nature:~a Chern insulator of non-interacting fermions and a fractional Chern insulator of hardcore bosons, both hosting a single chiral edge mode. From a practical point of view, our spectroscopic scheme should be readily implementable in cold atoms, only requiring  momentum-resolved (Raman) injection spectroscopy~\cite{Ness2020} applied to an impurity that is confined along the edge of the system.

More generally, this work contributes to an ongoing effort devoted to the characterization and detection of topological edge modes in ultracold atomic gases~\cite{goldman2010, stanescu2010, liu2010, goldman2012, goldman2013, reichl2014, stuhl2015, mancini2015,goldman2016, chalopin2020, braun2023,nardin2023,yao2024,elevator_2024,zhu2024}. In this context, previous works explored the spectroscopic characterization of chiral modes using well-designed probing fields~\cite{goldman2012,Binanti2024,nardin2024quantum}, quantized circular dichroism~\cite{unal2024}, and quench-dynamics protocols based on applying local perturbations on the edge of the sample~\cite{goldman2013,spielman2013,Dong2018, sun2023}. By injecting a mobile impurity locally on the edge of a topological system, and by characterizing the resulting spectroscopic response, the present work naturally contributes to this ongoing effort.

\subsection*{Outline}

The paper is structured as follows. In Sec.~\ref{sec:general}, we review the general polaron framework and define the impurity spectral function.  Moreover, we discuss the state-of-the-art of the main theoretical and numerical methods that are used in this context, with a special attention to tensor-network approaches. 
In Sec.~\ref{sec:1DFermi}, we study the Fermi polaron spectrum in the case of a non-chiral 1D lattice, demonstrating the accuracy of tensor-network methods and highlighting the lattice effects. The key features of the chiral polaron are investigated in Sec.~\ref{sec:1D_chiral}, using an effective 1D chiral model for the bath. The particle-hole excitation dynamics, as well as the lattice effects are also studied in that Section.
In Sec.~\ref{sec:Haldane}, we eventually study the formation of a chiral polaron on the edge of a realistic 2D topological (Chern) insulator. Here, the bath consists of a completely filled Bloch band of non-interacting fermions in the 2D Haldane lattice model. It is demonstrated that the impurity spectrum  inherits the chiral features of the edge modes, provided that the Fermi level lies in the topological gap and that the impurity is confined on the edge of the system. The power of tensor-network methods is then fully exploited in Sec.~\ref{sec:fractional}, where the chiral polaron emerges from the interaction with a (strongly-interacting) fractional Chern insulator. Here, the bath consists of hardcore bosons in the Harper-Hofstadter model.
Finally, we draw our conclusions and comment on possible perspectives in Sec.~\ref{sec:conclusions}.

\section{General framework: \\ polaron spectroscopy and methods}
\label{sec:general}

In this work, we consider the general setting of a mobile impurity interacting with a bath of particles, as described by a Hamiltonian of the form
\begin{equation}\label{eq:generalized_system_Hamiltonian}
    \hat{H} = \hat{H}_{\mathrm{bath}} + \hat{H}_{\mathrm{imp}} + \hat{H}_{\mathrm{int}}.
\end{equation}
Here, $\hat{H}_{\mathrm{bath}} $ denotes the many-body Hamiltonian describing the bath particles (fermions or bosons) moving on a lattice; its associated band structure can be trivial or topological, and this Hamiltonian may also include inter-particle interactions within the bath.

The operator $\hat{H}_{\mathrm{imp}}$ describes the  hopping of the mobile impurity on a subset $\mathcal{E} \subseteq \mathcal{L}$ of the lattice sites $\mathcal{L}$. For instance, in this work, we will consider restricting the impurity on the edge of a 2D lattice. It can be written as
\begin{align}\label{eq:generalized_impurity_Hamiltonian}
    \hat{H}_{\mathrm{imp}} &= - J_{\mathrm{I}} \sum_{\langle i, j\rangle} (\hat{d}_{i}^\dagger \hat{d}_{j}{+ \, {\rm h.c.}}), 
\end{align}
where $J_{\mathrm{I}}$ is the impurity hopping strength, 
$\braket{i,j}$ stands for a pair of nearest-neighboring sites, 
and $\hat{d}_{j}^\dagger$ denotes the impurity creation operator at lattice site $j \in \mathcal{E}$. Since we consider at most one impurity in the bath, our results are independent of its quantum statistics.

Finally, $\hat{H}_{\mathrm{int}}$ describes the interaction between the impurity and the bath particles. Considering onsite contact interactions between the two species, it can be written as
\begin{equation}\label{eq:generalized_int_Hamiltonian}
    \hat{H}_{\mathrm{int}} =
    U \sum_{j \in \mathcal{E}} 
    \hat{c}_{j}^\dagger
    \hat{c}_{j} \hat{d}_{j}^\dagger \hat{d}_{j},
\end{equation}
where $\hat{c}_{j}^\dagger$ denotes the creation operator of a bath particle and $U$ quantifies the interaction strength.

While ground-state properties of the coupled impurity-bath system will be analyzed below, we note that the most natural observable (in both ultracold atoms~\cite{knap2012} and solid-state experiments~\cite{sidler2017fermi}) is provided by the spectral function of the impurity. Our main target will thus concern the computation of impurity spectra, which can be accessed via injection spectroscopy in cold-atom setups. Experimentally, this corresponds to the following protocol: One starts from a state in which the impurity and the bath atoms do not interact (or only weakly interact). Then, the internal state of the impurity is resonantly flipped to a hyperfine level characterized by a sizable impurity-bath interaction. By utilizing Raman transitions, one can impart a finite momentum to the impurity, which makes it possible to map out its full momentum-resolved single-particle spectral function, corresponding to the final observable of the resonant excitation~\cite{Diessel_2022}.

Mathematically, the polaron spectral function is defined as
\begin{equation}
    \mathcal{A}(Q,\omega) =
    -2{\rm Im}
    \langle \Psi_0 | \hat{d}_{Q}
    \frac{1}{\omega - \hat{H} + E_{\mathrm{bath}}^0 + i\eta}
    \hat{d}^\dagger_{Q}
    | \Psi_0 \rangle,
    \label{eq:Aw}
\end{equation}
where $|\Psi_0\rangle$ is the ground state of the bath Hamiltonian $\hat{H}_{\mathrm{bath}}$ with energy $E_{\mathrm{bath}}^0$, while $\hat{H}$ is the full Hamiltonian including the interaction between the bath particles and the impurity. The spectral lines are artificially broadened by the linewidth $\eta$, describing effects such as Fourier broadening of the experimental injection pulse or a finite lifetime of the impurity. We assume that the impurity is both injected and detected in the quantum state labeled by $Q$, which will typically correspond to a momentum quantum number in the present context. We remark that the spectral function~\eqref{eq:Aw} satisfies the normalization 
$\int_{-\infty}^\infty \text{d}\omega ~ \mathcal{A}(Q,\omega)/ 2\pi \!=\! 1$
for each value of $Q$.
Furthermore, introducing the eigenstates and eigenenergies of the interacting Hamiltonian, 
$\hat{H} |n \rangle \!=\! E_{\rm int}^n |n \rangle$, the spectral function decomposes as
\begin{equation}
    \mathcal{A}(Q,\omega) =
    -2{\rm Im} \sum_{n}
    \frac{|\langle n | \hat{d}^\dagger_{Q}
    | \Psi_0 \rangle |^2}{\omega - E_{\rm int}^n + E_{\mathrm{bath}}^0 + i\eta},
    \label{eq:Aw_En}
\end{equation}
where the overlap 
$|\langle n | d^\dagger_{Q}| \Psi_0 \rangle |^2$ 
is usually referred to as the  oscillator strength of the state $|n\rangle$. In other words, the many-body states with a large (small) overlap with the injected state will be bright (dark), respectively. Unless otherwise stated, we set $\hbar \!=\! 1$ throughout this work.

\subsection*{Methods: \\ From the Chevy ansatz to matrix product states} \label{sec:general_MPS}

In the case of a fermionic bath, useful physical insight can be obtained by restricting the problem to the subspace associated with single particle-hole excitations on top of the bath's ground state:
\begin{equation}
    \hat{\mathcal{H}}_{\rm Chevy}
    =
    \left\{
    \hat{d}^\dagger_i |\Psi_0\rangle,
    \hat{d}^\dagger_i \hat{c}^\dagger_j \hat{c}_k |\Psi_0\rangle
    \right\}_{i\in\mathcal{E},jk\in\mathcal{L}}.
\end{equation}
This method was pioneered by Chevy \cite{chevy2006universal} for an impurity in a non-interacting Fermi sea, for which it proved to be a surprisingly accurate  approximation due to the destructive interference of higher-order terms with two or more particle-hole excitations~\cite{prokofev2008bold, combescot2008, vanhoucke2020}. It was recently adapted to investigate a number of different scenarios involving interacting fermionic backgrounds~\cite{Nishida2015,yi2015polarons,Pierce2019,Alhyder2022,amelio2023two-dimensional,mazza2022strongly,amelio2023polaron}.

However, for strongly-correlated environments, more sophisticated numerical methods are generally required, independently of the presence of the impurity. To begin with, Quantum Monte Carlo (QMC) methods have proved useful for studying the ground state of Bose polarons in dipolar atomic clouds~\cite{ardila2015impurity, ardila2020strong, sanchezbaena2023universal}, as well as weakly-interacting Fermi polarons~\cite{Prokofev_2008, bombin2019two-dimensional}. Obtaining spectral information with QMC methods is more challenging, as this requires ill-conditioned analytical continuation to real frequencies. Moreover, because of potential sign problems, correlated and/or topological backgrounds can be very challenging as well.

Then, it was recently shown in Ref.~\cite{amelio2024edpolaron} that exact diagonalization (ED) of small systems can provide valuable insights on the polaron spectra. In this context, the spectral function can be obtained using Krylov space methods. While ED will be used in the following to provide benchmarks on small systems, ED methods become prohibitively expensive as the size of the system is increased.

In this work, we use tensor-network methods as an unbiased numerical tool for computing polaron spectral functions. In particular, we employ the variational class of matrix product states (MPS). Originally, MPS were used as variational states for approximating ground states of correlated 1D quantum lattice systems, as in the density matrix renormalization group (DMRG) algorithm \cite{White1992, Schollwoeck2011}. The approximation can be systematically improved by increasing the bond dimension of the MPS. MPS methods have become so efficient that they are now also used for simulating 2D lattice systems on geometries such as cylinders or strips; for that purpose, the 2D lattice is mapped to a chain by snaking through the system along a 1D path. As a result, local 2D Hamiltonians are mapped to effective 1D Hamiltonians with longer-range terms, which makes the simulation substantially more costly and necessitates an exponential scaling of the MPS bond dimension.

Building on these ground-state algorithms, different approaches were developed for also accessing the low-energy excited states and spectral functions. Nowadays, the most efficient approach for computing a spectral function is through real-time evolution \cite{White2008}. In the present context of polaron physics, this amounts to first optimizing a ground-state approximation for the bath (either using the DMRG \cite{White1992} algorithm for a finite system, or the iDMRG \cite{McCulloch2008} or VUMPS \cite{ZaunerStauber2018} algorithms for infinite systems), and applying a local impurity creation operator on a site $j$ in the system
\begin{equation}\label{eq:general_impurity_creation_at_site_over_GS}
    \ket{\Psi_j(t=0)} = \hat{d}_j^\dagger \ket{\Psi_0}.
\end{equation}
This state can now be evolved in real time with the full bath-impurity Hamiltonian, where the time evolved state $\ket{\Psi_j(t)}\!=\!e^{-i\hat{H}t}\ket{\Psi_j(0)}$ is always approximated as an MPS. Early algorithms for MPS time evolution \cite{Vidal2004, Daley2004, White2004} relied on the Trotter-Suzuki decomposition, and are therefore unwieldy for longer-range Hamiltonians -- for instance, Hamiltonians involving non-local interactions or higher-neighbor hopping terms. Fortunately, time evolution based on the time-dependent variational principle (TDVP) \cite{Haegeman2011, Haegeman2016} or on matrix-product operator (MPO) representations of the time-evolution operator \cite{Zaletel2015, VanDamme2023} allow us to treat longer-range terms. 
After the time evolution, one can measure the dynamical correlation function, which we can Fourier transform to obtain the spectral function:
\begin{equation}\label{eq:general_spectral_fn_MPS}
    \mathcal{A}(Q,\omega) = \int_{-\infty}^\infty \mathrm{d}t \, e^{i\omega t} \sum_{j'} e^{iQ(j'-j)} \bra{\Psi_0} \hat{d}_{j'} \ket{\Psi_j(t)}.
\end{equation}
This method is essentially limited by two factors: (i) Because the entanglement entropy generically increases under unitary time evolution, the bond dimension in $\ket{\Psi_j(t)}$ would need to increase as a function of time; in practice, only a limited time window is accessible, which translates in a limited frequency resolution in the spectral function. (ii) The cylinder circumference or strip width is limited due to the exponential scaling of the bond dimension. 

While this more advanced methodology has allowed for precise evaluations of different types of Green's function and dynamic structure factors~\cite{Gohlke2017, Verresen2019, Bohrdt2020, Kadow2022, Xie2023, Sherman2023, Drescher2023}, for instance in the context of 2D quantum spin systems or Hubbard models, it has not been fully applied to the study of polaron formation yet. Indeed, only basic MPS methods -- such as time-evolving block decimation (TEBD) -- have been used for calculating polaron spectral functions, and were limited to 1D systems~\cite{Leskinen2010, massel2013dynamics, kantian2014, Visuri2016, Visuri2017, Kamar2019}. In this work, we show that an extension of state-of-the-art methods can be used to tackle 2D systems as well; see Appendix~\ref{sec:MPS_simulations} for details. We will apply this methodology to compute the dynamics of impurities immersed in topological 2D systems, with and without (strong) interactions in the bath, both in the bulk and on the edge of the system.

\section{Fermi polaron on a 1D lattice:\\ the non-chiral case}
\label{sec:1DFermi}

Before addressing the chiral polaron formation in topological systems, it is important to build a clear understanding of the impurity dynamics in a 1D, non-interacting, Fermi sea without any time reversal symmetry breaking.
Here, we focus on lattice systems, motivated by the availability of powerful, quasi-exact, numerical methods (such as ED and MPS) in this framework. As we will see, some specific features of the impurity spectra arise from the 1D and lattice nature of the system. With respect to the existing literature, an impurity in a 1D Fermi gas has been the subject of a few analytical and numerical works~\cite{rosch1995heavy,lamacraft2009dispersion,Leskinen2010,Guan2011,Mathy2012,kantian2014,liu2020nongaussian}, both on a lattice and in the continuum (where an effective Tomonaga-Luttinger liquid description is typically assumed for the background). However, to the best of our knowledge, a full frequency- and momentum-resolved impurity spectrum has not been computed using quasi-exact numerical methods, and a detailed account of lattice effects has remained elusive. For these reasons, we devote this section to presenting a self-contained description of the momentum-resolved spectrum of an impurity injected in a 1D lattice Fermi sea.

The Hamiltonian of an impurity immersed in a non-interacting Fermi sea, on a 1D lattice, takes the form of Eq.~\eqref{eq:generalized_system_Hamiltonian}, with the impurity and interaction terms given by Eqs.~\eqref{eq:generalized_impurity_Hamiltonian}-\eqref{eq:generalized_int_Hamiltonian} and
\begin{equation}\label{eq:system_hamiltonian_non-chiral}
\hat{H}_{\mathrm{bath}} = \hat{H}_{\mathrm{1D}} = -J\sum_{j=0}^{L-1} \left(\hat{c}^\dagger_j \hat{c}_{j+1} + \hat{c}^\dagger_{j+1} \hat{c}_j \right),
\end{equation}
where $J$ is the nearest-neighbor hopping strength of bath particles. In the following, we assume periodic boundary conditions, $j\!=\!0,1,..,L-1$, and consider $N$ spinless fermions in the system. In momentum space, the Hamiltonians  of the bath  fermions and of the impurity read, respectively,
\begin{subequations}
\begin{align}
    \hat H_{\mathrm{bath}}
    &= \sum_k\varepsilon_\text{bath}(k)\hat c_k^\dagger\hat c_k 
    =-2J\sum_k\cos (ka)\hat c_k^\dagger\hat c_k,  
    \\
    \hat H_{\mathrm{imp}}
    &= \sum_k\varepsilon_\text{I}(k)\hat d_k^\dagger\hat d_k 
    =-2J_{\mathrm{I}}\sum_k \left(\cos (ka) - 1 \right) d_k^\dagger\hat d_k, 
    \label{eq:impurityenergy}
\end{align}
\end{subequations}
with $k\!=\!0,1,...,(L-1) \times 2\pi/La$ the quasi-momenta defined modulo $2\pi/a$, and where $a$ is the lattice constant. We have added the constant $2J_{\mathrm{I}}$  to the impurity dispersion $\varepsilon_\text{I}(k)$, such that its band minimum has zero energy. If $N$ is odd, the ground state of the fermionic bath (without the impurity) is simply given by the non-degenerate Fermi sea state
\begin{equation}
|\Psi_0\rangle \equiv |\mathrm{FS}\rangle = \prod_{|k| < k_F} \hat{c}^\dagger_k |0\rangle, 
\end{equation}
where $k_F\!=\!\pi N / La$ is the Fermi momentum.

In the following, we will investigate the energy- and momentum-resolved impurity spectral function $\mathcal{A}(Q,\omega)$ using three different methods: ED, MPS and the Chevy ansatz. In the ED case, we will use Krylov space methods to exactly calculate the impurity spectral function for small system sizes, yet presenting reasonably weak finite-size effects~\cite{amelio2024edpolaron}.  Next, we will use the ED spectra to benchmark the spectral function obtained using the computationally cheaper MPS and Chevy-ansatz methods. 

\begin{figure*}[t]
\centering
\includegraphics[width=\textwidth]{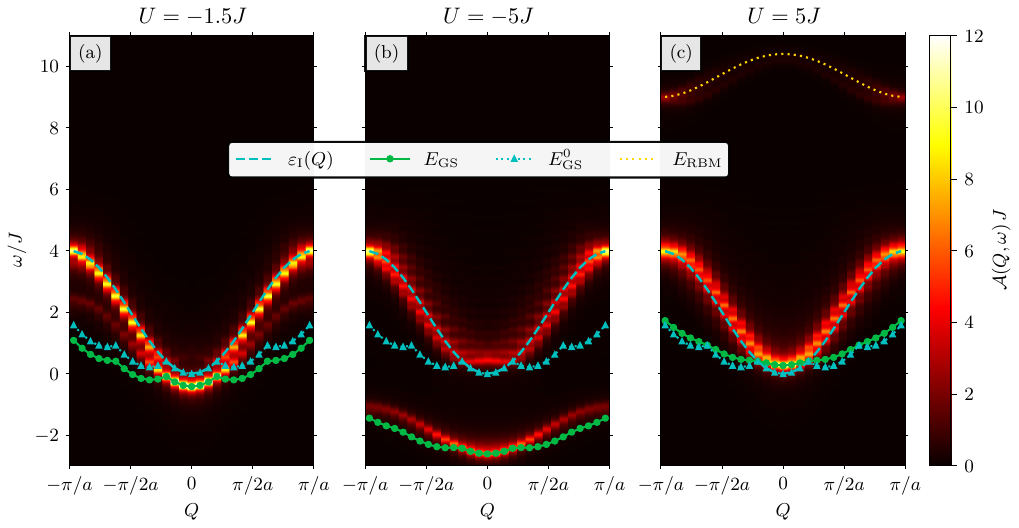}  
\caption{
Spectral function $\mathcal{A}(Q,\omega)$ of an impurity injected in a 1D non-interacting Fermi sea, as calculated using ED. The three panels correspond to different impurity-fermion interaction strengths:~(a) $U\!=\!-1.5J$, (b) $U\!=\!-5J$, and (c) $U\!=\!5J$. We fixed $J_{\mathrm{I}}=J$, and considered $N\!=\!5$ fermions in a 1D lattice with $L\!=\!29$ sites and periodic boundary conditions. The dashed cyan line indicates the bare impurity dispersion $\varepsilon_{\mathrm{I}}(Q)$, the green circles correspond to the many-body ground state energy $E_{\mathrm{GS}}$ at the given $U$, while the cyan triangles denote the many-body ground state energy $E_{\mathrm{GS}}^0$ at $U\!=\!0$; the yellow dots in panel (c) are the energies $E_{\mathrm{RBM}}$ of the two-body repulsively bound state. The lower energy peak in (a) and (b) at a given $Q$ corresponds to the attractive polaron (AP) and the higher energy one to the repulsive polaron (RP).
}
\label{fig:trittic_1DFermi_ED}
\end{figure*}

\subsection{ED results}

The key features of the spectral function for an impurity interacting with the 1D Fermi sea are illustrated in Fig.~\ref{fig:trittic_1DFermi_ED}, where we plot  $\mathcal{A}(Q,\omega)$ obtained with ED on a ring geometry (periodic boundary condition) with $L\!=\!29$, $N\!=\!5$ and $J_{\mathrm{I}}\!=\!J$. The three panels correspond to different impurity-bath interaction strengths, namely weak attraction $U\!=\!-1.5J$ (a), strong attraction $U\!=\!-5J$ (b) and  strong repulsion $U\!=\!5J$ (c). The bare impurity dispersion $\varepsilon_{\mathrm{I}}(Q)$ is indicated by the cyan dashed lines.

In panels (a) and (b), one notices a sharp lower branch at small momenta and a bright upper branch at large momenta, which progressively split as one increases the interaction strength $|U|$; we will refer to these as the attractive (AP) and repulsive (RP) polaron branches, respectively. Interestingly, the RP shows a negative effective mass at small momenta, whereas the RP peak is very narrow and close to the bare dispersion for $Q \sim \pi/a$. Instead, in panel (c), one observes that the bare impurity dispersion is slightly blue-shifted compared to the non-interacting case, and that it broadens at momenta $|Q| \sim \pi/2a$. Furthermore, a high-energy branch appears (barely visible in this specific plot), which we ascribe to the formation of a repulsively bound molecule. Indeed, its position is captured by the energy $E_{\mathrm{RBM}}$ obtained by solving the two-body problem of one fermion interacting with one impurity; see the yellow dots in panel (c).

In Fig.~\ref{fig:trittic_1DFermi_ED}, we also plot the many-body ground-state energy $E_{\mathrm{GS}}(Q)$ of the coupled impurity-bath system, for each $Q$ sector and for a single impurity particle; see the green circles. The first peculiarity concerns the oscillatory behavior of $E_{\rm GS}(Q)$. In fact, this effect was predicted in Ref.~\cite{lamacraft2009dispersion}, and it is explained as follows:~in the limit $L\to\infty$, Umklapp processes  can create a particle-hole excitation of momentum $2k_F$ across the Fermi sea at a negligible energy cost. As a consequence, $E_{\mathrm{GS}}(Q)$ has to be periodic in $2k_F$ in the thermodynamic limit. In a finite-size system, creating such an excitation costs an energy of order $1/L$, which lifts the degeneracy. This is also illustrated by the cyan triangles, which represent $E^0_{\mathrm{GS}}(Q)$, the energy of the many-body ground state at $U\to0$. The second interesting feature is that the ground state is mainly dark for momenta larger than $k_F$  i.e. $\mathcal{A}(Q, E_{\mathrm{GS}})\simeq 0$. This is due to the fact that the ground state at $|Q|>k_F$ consists of an impurity with small momentum plus a few Umklapp particle-hole excitations. These states have negligible overlap with the non-interacting impurity at $Q$, and are thus dark according to Eq.~\eqref{eq:Aw_En}.

\subsection{Benchmarking MPS and Chevy methods}

\begin{figure*}[t]
\centering
\includegraphics[width=\textwidth]{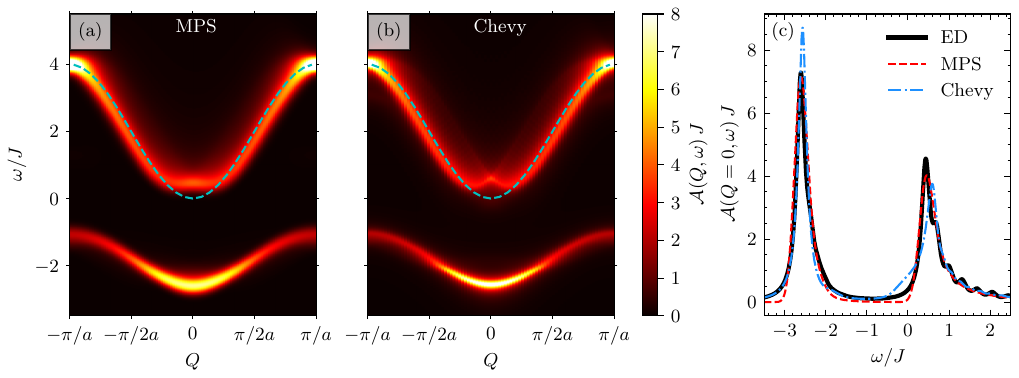}  
\caption{
Benchmarking the MPS and Chevy ansatz. Spectral function $\mathcal{A}(Q,\omega)$ of an impurity injected in a 1D Fermi sea, as computed using: (a) MPS, and (b) Chevy ansatz calculations. Here we set $U\!=\!-5J$ and considered the density $N/L\!=\!5/29$, for both the MPS and Chevy calculations. The dashed cyan line indicates the bare impurity dispersion $\varepsilon_I(Q)$. In panel (c) we compare the impurity spectrum at $Q\!=\!0$ for ED, MPS and the Chevy ansatz. In the ED and MPS results, there is a slight asymmetry of the AP peak that hints at the absence of quasi-particle behavior. For the MPS calculation, we used a bath of $N\!=\!25$ fermions in a finite chain of length $L\!=\!145$, the maximum bond dimension was fixed at 150 with an SVD cutoff $10^{-12}$ in the DMRG algorithm used to find an approximation to $| \mathrm{FS} \rangle$, and we performed time-evolution using TDVP in time steps of length $dt\!=\!0.2 / J$ with an SVD cutoff $10^{-8}$, terminating it at a time $t_f\!=\!25 /J$.
}
\label{fig:trittic_1DFermi_comparison}
\end{figure*}

We now build on these results, which are exact (although for a relatively small system), to benchmark the calculation of the spectral function obtained from MPS time-evolution methods and the variational Chevy ansatz, for a fixed value of the interaction constant $U\!=\!-5J$ and density $N/L\!=\!5/29$.

First, we show the MPS results in Fig.~\ref{fig:trittic_1DFermi_comparison}(a), which are calculated using the general procedure described in Sec.~\ref{sec:general_MPS}. By comparing with the ED results, we notice that the MPS method gives an excellent approximation to the translationally-invariant ED calculation performed above keeping the density $N/L$ fixed; this is further illustrated in Fig.~\ref{fig:trittic_1DFermi_comparison}(c). The discrepancy in the spectral function can be attributed to the differing artificial broadening of the spectral peaks introduced in the calculation of spectral functions:~a Lorentzian broadening was utilized for ED, while a Gaussian one was used for MPS calculation. However, more importantly, the AP and RP peak positions, and their spectral ratios are in excellent agreement for the two cases.

Fig.~\ref{fig:trittic_1DFermi_comparison}(b) shows the spectral function obtained from the Chevy ansatz
\begin{equation}\label{eq:chevy_ansatz_1D}
    |\Psi^Q \rangle = \left(\psi^Q \hat{d}^\dagger_Q
    +
    \sum_{kp}{}^{'} \psi_{kp}^Q\, \hat{d}^\dagger_{Q-k+p} \hat{c}^\dagger_{k} \hat{c}_{p} \right) |\mathrm{FS}\rangle.
\end{equation}
Here, $\psi^Q$ and $\psi_{kp}^Q$ are the variational parameters and primed sums indicate that the summation occurs over momenta fulfilling $k>k_F$ and $p<k_F$. The label $Q$ indicates that - thanks to translational invariance - each total momentum sector is treated separately and is determined by the momentum at which the impurity is injected.
The Schr\"odinger equations in the Chevy subspace read
\begin{subequations}
\begin{small}
\begin{align}
    E \psi^Q
    &=
   \left( \varepsilon_{\mathrm{I}}(Q) + \frac{UN}{L}
   \right)
    \psi^Q
    +
   U \sum_{kp}{}^{'}
    \psi_{kp}^Q , \\
    E \psi^Q_{kp}
    &=
    \left(
    \varepsilon_{\rm bath}(k) - \varepsilon_{\rm bath}(p) +
    \varepsilon_{\rm I}({Q-k+p})
    +
    \frac{UN}{L}
    \right)
    \psi_{kp}^Q
    + \nonumber \\
    & \quad +
       U \psi^Q
    + U \sum_{q} (\psi_{k+q,p}^Q - \psi_{k,p+q}^Q).
\end{align}
\end{small}
\end{subequations}
These equations are diagonalized numerically for $N\!=\!15$ particles and $L\!=\!87$ sites, corresponding to the same density as for the ED calculation. The spectral function is then obtained from Eq.~\eqref{eq:Aw_En}.

Overall, we find that the Chevy approximation accurately describes the spectral function of the impurity. This might appear surprising, since it is known that the quasi-particle picture breaks down in 1D, and an orthogonality-type catastrophe occurs~\cite{rosch1995heavy,kantian2014}. This is indeed confirmed in panel (c), where the spectral function obtained from ED, MPS, and the Chevy ansatz are plotted for $Q\!=\!0$. The AP peak obtained through the Chevy ansatz exhibits a narrow Lorentzian peak corresponding to a well-defined quasi-particle, and a small detached molecule-hole continuum around $\omega \simeq -2J$. Instead, in the ED and MPS cases, we find an asymmetric peak, suggesting a power law decay on the high-energy side, corresponding to a break-down of the quasi-particle picture. The peak is, however, quite narrow and well described by the Chevy-polaron ansatz as well as by the MPS. 

From this we conclude that, even though the Chevy ansatz does not capture the conceptually important absence of quasi-particles in 1D, from a practical level, it does provide a very good estimation of how the impurity spectra would appear in experiments, where extrinsic broadening of the lines is present and where resolving the small discrepancies highlighted in panel (c) would be very challenging. 

It is also useful to compare these results with the 2D results reported in Fig.~7 of Ref.~\cite{amelio2024edpolaron}, where all secondary peaks are well detached from the Lorentzian AP peak, in both ED and Chevy calculations. Concerning the RP branch, one can state that there is excellent agreement between the ED and Chevy results, and that only very small differences are visible.

\section{Impurity in a 1D chiral bath: the chiral polaron}\label{sec:1D_chiral}

We now address the main topic of this work:~the properties of a mobile impurity interacting with a chiral many-body environment. 

Chiral modes naturally arise on the edge of 2D systems with broken time-reversal symmetry, such as quantum Hall states or (fractional) Chern insulators. In general, these systems are incompressible in the bulk (the ground state is gapped on the torus), they are characterized by a quantized Hall conductivity (i.e.~a non-zero many-body Chern number), and they display gapless chiral edge modes~\cite{qi2011}. These systems can be realized in various ways, e.g.~by subjecting a 2D material to an external magnetic field, or by engineering a synthetic gauge field in an ultracold atomic gas~\cite{aidelsburger2018}.

In principle, when immersing a mobile impurity in a quantum-Hall-type background, one has to consider the dressing of the impurity by the bulk modes as well as by the edge modes. In a quantum Hall or Chern insulator state, the Fermi energy falls within a bulk gap, such that the bulk modes are gapped while the edge modes are gapless. Here, we assume that the impurity is confined on the edge of the system, such that the coupling to the bulk modes is strongly reduced, both for energetic reasons and due to the small overlap of bulk modes with the edge. The validity of this assumption will be analyzed in Sec.~\ref{sec:Haldane}.

In this Section, we aim at elucidating the key properties and signatures associated with the dressing of a mobile impurity by the low-energy chiral edge excitations, neglecting the effects of bulk modes. To do so, we will introduce a simple toy model describing a single mobile impurity, characterized by a trivial quadratic dispersion, interacting with a Fermi sea of chiral, linearly dispersing fermions. In particular, this simplified setting reveals how the resulting dressing induces characteristic chiral features in the impurity spectral function.

\subsection{Phenomenological 1D chiral model}

We model the chiral edge mode with a linear spectrum $\varepsilon_{\mathrm{bath}}(k)\!=\!vk$, where $v$ is the group velocity. In a 2D quantum-Hall-type system, edge states are located within an energy gap $\Delta_{\rm gap}$, and thus, one can assume that they have momenta $k \in [-\Lambda/2, \Lambda/2]$, where the cutoff momentum $ \Lambda\!=\!\Delta_{\rm gap} / v$ also defines a typical length scale $\ell\!=\!2\pi/\Lambda$. We point out that shifting this momentum interval by a constant does not have any physical consequence, as it is a gauge choice. 

In a conventional integer quantum Hall system (i.e.~non-interacting electrons subjected to a uniform magnetic field in the continuum),  $\Delta_{\rm gap}$ corresponds to the cyclotron frequency $\omega_B\!=\!eB/m$ and the group velocity is proportional to the slope of the confining potential at the edge; here $e$ and $m$ denote the charge and mass of the electron, and $B$ is the external magnetic field. For a sharp edge, the typical length $\ell$ is of the order of the magnetic length 
$\ell_B\!=\!\sqrt{\hbar/eB}$. 
In the case of a Chern insulator (e.g.~non-interacting fermions defined on the Haldane lattice model), the gap is of the order of the hopping strength, and $\ell$ is of the order of the lattice constant; see also Sec.~\ref{sec:Haldane}.

Based on these observations, we introduce a simplified toy-model Hamiltonian
\begin{subequations}\label{eq_chiral1D}
\begin{align}
    \hat{H}_{\mathrm{bath}} &\equiv \hat{H}_{\mathrm{chiral}}
     = \sum_k [ \varepsilon_{\mathrm{bath}}(k) -\mu  ] \hat{c}_k^\dagger \hat{c}_k, \\
    \hat{H}_{\mathrm{imp}} &= \sum_q \varepsilon_{\mathrm{I}}(q) \hat{d}_q^\dagger \hat{d}_q, \\
    \hat{H}_{\mathrm{int}} &= \frac{1}{N_P} \sum_{kpq}
         U_{kp} \hat{c}_k^\dagger \hat{c}_p \hat{d}^\dagger_{q+p-k} \hat{d}_q,
\end{align}
\end{subequations}
where the fermionic momentum  is discretized in $N_{P}$ states,   $k\!=\!-\frac{\Lambda}{2} +  m \frac{\Lambda}{N_P+1}$, with $m\!=\!1,...,N_{P}$, such that  $k \in \left(-\frac{\Lambda}{2},\frac{\Lambda}{2}\right)$. The dispersion of the impurity is assumed to be parabolic $ \varepsilon_{\mathrm{I}}(q)\!=\!q^2 /2M $, with mass $M$ and momentum $q \in (-\infty,+\infty)$.

The chemical potential $\mu$ sets the number of occupied edge states in the non-interacting ground state. Taking $\mu\!=\!0$  corresponds to half filling, with $N \!=\! N_P/2$ fermionic states occupied (assuming $N_P$ even). The corresponding ground state, without the impurity, is a chiral Fermi sea 
\begin{equation}
    \ |\Psi_0\rangle \equiv |\mathrm{FS}\rangle = \prod_{k<0} \hat{c}^\dagger_k |0\rangle.
\end{equation}
The interaction term $\hat{H}_{\mathrm{int}}$ conserves the total momentum, defined as
\begin{equation}
\hat{Q} = \sum_k k \ \hat{c}_k^\dagger \hat{c}_k
    +
    \sum_q q \ \hat{d}_q^\dagger \hat{d}_q
    - \langle \mathrm{FS}| \sum_k k \ \hat{c}_k^\dagger \hat{c}_k | \mathrm{FS} \rangle, 
\end{equation}
where we explicitly subtracted a constant such that the total momentum coincides with the momentum at which the impurity is injected in the Fermi sea. In principle, realistic interaction terms should take the wavefunctions  of the chiral edge states into account, through the coefficients $U_{kp}$ in Eq.~\eqref{eq_chiral1D}. However, since this is not necessary to capture the essential features of the chiral polaron formation, we simply set $U_{kp}=U$, i.e.~a constant, in the rest of this Section.

\subsection{Chevy ansatz analysis}

We calculate the single-particle spectral function of the impurity by diagonalizing the problem in the truncated Hilbert space provided by the Chevy ansatz, see Sec.~\ref{sec:general}. This allows one to gain valuable physical insight into the formation of the polaron through single particle-hole excitations of the chiral bath 
\footnote{In two- and three-dimensions, a UV-cutoff $\Lambda$ is required to regularize the theory where the term $\frac{1}{N_P}\sum_{q'}{}^{'}U_{q'\!q}\psi_{kq}^Q$ drops due to regularization when we write $U$ using the Lippmann-Schwinger equation $\frac{1}{g}=\frac{\mu}{2\pi a}-\frac{1}{V}\sum_{p}\frac{1}{\varepsilon^c_p+\varepsilon^d_p}$. However, in one dimension, a UV cutoff is not required, and thus, we find a polaron in one dimension even in the case of microscopic repulsive interactions.}.
Momentum conservation is used to treat each total momentum sector $Q$ independently. Here, it is particularly convenient to measure energies in units of $Mv^2$ and momenta in units of $Mv$, which corresponds, respectively, to the kinetic energy and momentum of the impurity when moving at the group velocity of the chiral bath. Hence, this leaves us with only three parameters:~the total momentum $Q/Mv$, the interaction strength $U/Mv^2$, and the  cutoff momentum $\Lambda/Mv$ of the chiral modes. In principle, one does not need a momentum cutoff in 1D, since the energy is well-defined in the limit  $\Lambda \to \infty$. We shall nevertheless retain a finite $\Lambda$ in our numerical calculations, motivated by the fact that such a cutoff has a physical origin in the relevant case of a 2D Chern-insulator, where it is set by the size of the topological bulk gap. Also, we will study the effects of particle-hole symmetry breaking, which occurs for a bounded edge mode (i.e.~for finite $\Lambda$) whenever $\mu \neq 0$.

\begin{figure}[t]
\centering
\includegraphics{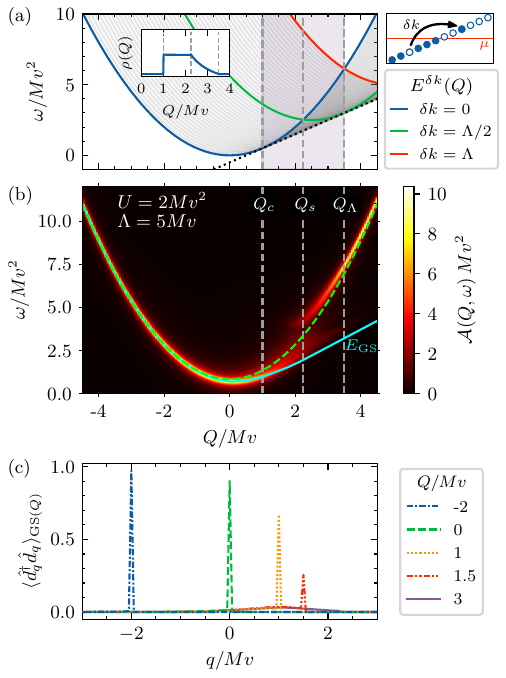} 
\caption{
(a) Non-interacting spectrum of the system from Eq.~\eqref{eq:non_interacting_parabolic_spectrum}, as a function of the injected momentum   $Q$ and for different values of $\delta k \in [0,\Lambda]$. The vertical dashed lines correspond, from left to right, to $Q_c, Q_s, Q_\Lambda$; the pink shaded area indicates the existence of particle-hole states resonant with the bare impurity (blue curve). Inset: density of  particle-hole states resonant with the bare impurity.
(b) Spectral function  $\mathcal{A}(Q,\omega)$ of an impurity in a chiral bath. The dashed green line indicates the bare impurity dispersion (including the mean-field shift $U/2$), while the cyan line denotes the  energy $E_{\rm GS}(Q)$ of the ground state, which becomes gradually darker for $Q>Q_c$.
(c) Momentum distribution $\langle \mathrm{GS}(Q)| \hat{d}^\dagger_q \hat{d}_q |\mathrm{GS}(Q)\rangle$ of the impurity in the ground state, illustrating the threshold behavior at $Q_c$. Different colors indicate different momentum sectors $Q$. 
The parameters used are $U/Mv^2\!=\!2$, $\Lambda \!=\! 5Mv$, $N_P\!=\!100$.
}
\label{fig:trittic_1D_chiral}
\end{figure}

We start by analyzing the spectrum of the non-interacting ($U=0$) impurity-bath system, considering single particle-hole excitations in the chiral bath. Calling $\delta k$ the momentum of such a particle-hole excitation [see the sketch in the inset of Fig.~\ref{fig:trittic_1D_chiral}(a)], the non-interacting spectrum is given by
\begin{equation}
\label{eq:non_interacting_parabolic_spectrum}
    E^{\delta k}(Q) = \varepsilon_{\mathrm{I}}(Q-\delta k) + \varepsilon_{\mathrm{bath}}(\delta k)= \frac{(Q- \delta k)^2}{2M} + v \delta k.
\end{equation}
This family of curves is represented by thin gray curves in Fig.~\ref{fig:trittic_1D_chiral}(a), as a function of the momentum of the injected impurity $Q$ and parameterized by the transferred momentum $\delta k \in [0, \Lambda]$. Geometrically, the different (gray) parabolas share a common tangent of slope $v$ (thick black dots).

Three particularly relevant curves are highlighted by the thick blue, green and red lines. The blue curve is the bare impurity dispersion in the absence of particle-hole excitations, or, in other terms, at $\delta k =0$.
The red curve corresponds to the maximum momentum transfer $\delta k\!=\!\Lambda$, and it intersects the bare impurity dispersion at $Q_\Lambda\!=\!Mv + \Lambda/2$, satisfying $E^{0}(Q_\Lambda)\!=\!E^{\Lambda}(Q_\Lambda)$.
Finally, the green curve is obtained for  $\delta k\!=\!\Lambda/2$, and it meets the blue curve at $Q_s\!=\!Mv + \Lambda/4$. The importance of this green curve will become clear in the next subsection; at this stage, let us simply note that the expression for $Q_s$ provided above only holds at half-filling.

At $U\!=\!0$, the impurity spectral function is peaked at the bare impurity energy, which takes up all the oscillator strength. When the interaction $U$ is switched on, the non-interacting levels will hybridize and the oscillator strength will be distributed between different many-body states. This hybridization is enhanced by the existence of resonances  at the non-interacting level, i.e.~crossings between the blue and grey curves in Fig.~\ref{fig:trittic_1D_chiral}(a). This motivates us to inspect the density of single-particle-hole states at the bare impurity energy $Q^2/2M$, which we denote as  $\rho(Q)$; see Appendix \ref{app:DOS_chiral} for the precise definition and computation of $\rho(Q)$. This density of states, which takes the degeneracy of $E^{\delta k}(Q)$ into account, is plotted in the inset of Fig.~\ref{fig:trittic_1D_chiral}(a). Importantly, $\rho(Q)$ is found to be nonzero for $Q_c \leq Q \leq Q_\Lambda$, with a cusp at $Q\!=\!Q_s$; see also the vertical dashed lines and the pink shaded region in Fig.~\ref{fig:trittic_1D_chiral}(a). Here, the critical momentum is given by the intuitive relation 
\begin{equation}
Q_c\!=\!Mv.
\end{equation}
Indeed, it derives from the condition 
$\partial_{\delta k} E^{\delta k}(Q_c)|_{\delta k=0}\!=\!0$,
or equivalently, $\partial_Q \varepsilon_I(Q_c) \!=\! v$, which has a clear physical interpretation:~when the impurity is moving at momentum $Q_c\!=\!Mv$,  i.e.~when the velocity of the impurity matches the group velocity of the chiral background medium, the impurity can resonantly exchange energy-momentum with the bath.

Building on these analytical arguments, we now move to the most important result of this Section. In Fig.~\ref{fig:trittic_1D_chiral}(b), we present the impurity spectral function $\mathcal{A}(Q,\omega)$, obtained for an interaction strength $U/Mv^2\!=\!2$, at half-filling ($\mu=0$); the cutoff is set to $\Lambda\!=\!5Mv$. The most striking feature concerns the asymmetry of the spectral function, which is distinguished by a pronounced broadening of the impurity spectral peak in the range delimited by the critical momenta $Q_c$ and $Q_{\Lambda}$, and signals the formation of a polaron cloud. Moreover, a splitting is observed at $Q\!=\!Q_s$, with a  magnitude increasing with $U$. As previously explained, the interval $[Q_c, Q_\Lambda]$  corresponds to the existence of particle-hole states that are nearly resonant with the bare impurity dispersion (dashed green curve). On the contrary, when $Q<Q_c$ or $Q>Q_\Lambda$, the dressing is strongly suppressed, due to the fact that scattering processes are non-resonant; in this regime, the spectral function can thus be accurately determined from the (mean-field shifted) bare-impurity dispersion in Eq.~\eqref{eq:non_interacting_parabolic_spectrum}.

It is also insightful to study the ground state of the system as a function of $Q$. The ground state energy is plotted as a cyan line in Fig.~\ref{fig:trittic_1D_chiral}(b). For $Q \lesssim Q_c$, the ground state is bright, and its energy follows the bare impurity dispersion. In contrast, for $Q \gtrsim Q_c$, the ground-state energy disperses linearly according to the chiral-bath group velocity $v$, and the ground state becomes dark in the impurity spectral function. This linear dispersion, which can be simply interpreted as the impurity being dragged by the chiral bath at velocity $v$, can be obtained by minimizing the non-interacting spectrum $E^{\delta k}(Q)$ in Eq.~\eqref{eq:non_interacting_parabolic_spectrum} with respect to $\delta k$. This simple picture also explains the darkness of the ground state in the spectral function at large $Q\gg Q_c$:~injecting an impurity with a very high velocity $v_I \gg v$ into the system has vanishing overlap with the ground state of the coupled impurity-bath system, drifting at velocity $v$.
 
This picture is confirmed in Fig.~\ref{fig:trittic_1D_chiral}(c), where we plot the impurity-momentum distribution in the many-body ground state, $\langle \mathrm{GS}(Q)| \hat{d}^\dagger_q \hat{d}_q |\mathrm{GS}(Q)\rangle$, for five different values of the total momentum $Q$. One finds that the impurity-momentum distribution has a delta peak at the injected momentum, i.e.~at ~$q\!=\!Q$, with strength decreasing to zero as $Q$ increases; this signals a slowly vanishing overlap with the bare impurity and results in the ground state becoming darker. In addition, for $Q>Q_c$, the distribution develops a broader peak pinned at the critical momentum $q\!=\!Q_c\!=\!Mv$.

We demonstrate the validity of these results beyond the Chevy approximation by providing an ED benchmark in Appendix~\ref{app:ED}.

\subsection{Particle-hole duality}
\label{ssec:1D_chiral_PH_symmetry}

To gain more insight into the physics occurring between the critical points $Q_c$ and $Q_{\Lambda}$, and in particular around $Q_s$, we now analyze the contribution of particle and hole dynamics separately. Here, it will prove useful to consider imbalanced particle and hole phase spaces, namely, to tune the chemical potential away from half-filling (i.e.~$\nu\!\ne\!1/2$).

\begin{figure}[t]
\centering
\includegraphics{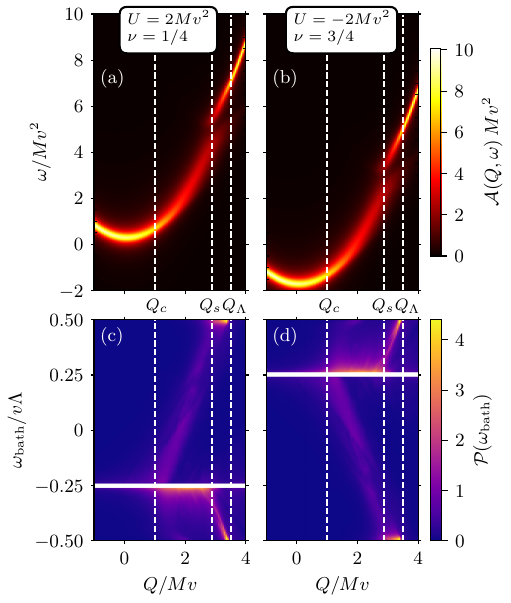}  
\caption{
Particle-hole duality. Spectral function $\mathcal{A}(Q,\omega)$ for: (a) $U \!=\!2Mv^2$, at filling $\nu\!=\!0.25$; and (b) $U\!=\!-2Mv^2$ at filling $\nu \!=\! 0.75$. The spectra are the same, but for a vertical shift of $U$, reflecting the particle-hole duality.  The bottom row (c-d) shows the occupation of particle and hole states at energy $\omega_{\rm bath}$, averaged over all the many-body eigenstates and weighted by their oscillator strength. The white line denotes the Fermi level. 
}
\label{fig:PHsymmetry}
\end{figure}

First, it is instructive to compare the impurity spectral function of a given impurity-bath configuration with that of its \emph{dual}, obtained by reversing the sign of the impurity-bath interaction $U$ and modifying the filling factor $\nu$ according to:
\begin{equation}
\nu \rightarrow 1-\nu , \qquad U \rightarrow - U.\label{duality_transf}
\end{equation}
As a concrete example, Fig.~\ref{fig:PHsymmetry}(a) displays the impurity spectral function for a filling factor $\nu\!=\!1/4$ and repulsive interactions $U\!=\!2Mv^2$, while Fig.~\ref{fig:PHsymmetry}(b) shows the {\em dual} case, with $\nu\!=\!3/4$ and attractive interactions $U\!=\!-2Mv^2$. In both cases, the bath cutoff is fixed to $\Lambda\!=\!5Mv$. Importantly, one notices that both spectra are identical, apart from the mean-field energy shift, which transforms according to $\nu U \to -U(1-\nu)$. Moreover, the splitting point $Q_s$ is shifted to the right with respect to the half-filling case shown in Fig.~\ref{fig:trittic_1D_chiral}(b), as we will explain below.

In order to elucidate the correspondence, or {\em duality}, between the two spectra of Fig.~\ref{fig:PHsymmetry}, we introduce the particle-hole transformation $\hat{h}_k = \hat{c}_{-k}^\dagger$, and we apply it to the Hamiltonian in Eq.~\eqref{eq_chiral1D}. Anti-commuting and redefining the fermionic momenta in the sums, the transformed Hamiltonian becomes
\begin{align}
    \hat{H} &= \sum_k \left(\varepsilon_{\rm bath}(k) + \mu \right) \left(1 + \hat{h}_k^\dagger \hat{h}_k \right) + \sum_q \varepsilon_{\rm I}(q) \hat{d}_q^\dagger \hat{d}_q \nonumber \\
    &\quad - \frac{1}{N_P} \left[ \sum_{kpq} U_{kp} \hat{h}_k^\dagger \hat{h}_p \hat{d}^\dagger_{q+p-k} \hat{d}_q + \sum_{k q} U_{kk}\, \hat{d}^\dagger_{q} \hat{d}_q\right].
    \label{eq:dualH}
\end{align}
This  differs from the original Hamiltonian through four features:~(i)~the sign of $\mu$; (ii)~an irrelevant constant, $\sum_k (\varepsilon_{\rm bath}(k) + \mu)$; (iii)~the sign of the interaction; (iv)~a shift of the impurity energy in the last term. We remark that this last feature reduces to a shift of $\frac{1}{N_P}\sum_{k} U_{kk}\!=\!U$ for a single impurity, if the interaction is momentum independent. Altogether, this demonstrates that the Hamiltonian at filling factor $\nu$ and interaction strength $U$ is equivalent to the particle-hole transformed Hamiltonian at filling factor $(1-\nu)$ and $-U$, plus a $U$ shift in the impurity energy. This result provides a simple explanation as to why Fig.~\ref{fig:PHsymmetry}(a) and Fig.~\ref{fig:PHsymmetry}(b) are directly related by an energy shift of $U$.

To investigate the individual roles of the bath particles and holes in the polaron formation, we analyze the occupation $\mathcal{P}(\omega_{\mathrm{bath}})$ of particle and hole states (labeled by their energy
$\omega_{\mathrm{bath}} \in [-v\Lambda/2, v\Lambda/2]$),
averaged over all the many-body eigenstates of the impurity-bath system and weighted by their oscillator strength, for each $Q$ sector. The results are shown in Fig.~\ref{fig:PHsymmetry}(c-d), for the two cases investigated above; the corresponding formulas are provided in Appendix \ref{subapp:p-h_probability}.

Let us start with the repulsive case, shown in panel Fig.~\ref{fig:PHsymmetry}(c). First of all, we note that the dressing is weak for $Q < Q_c\!=\!Mv$, since, as previously discussed, all the bare (single) particle-hole states are off-resonant with respect to the injected impurity. Increasing $Q$ beyond $Q_c$, one finds that the cloud of holes that dresses the impurity is broad and peaks at the Fermi level, whereas the particle occupancy peak shifts with the excess momentum $Q\!-\!Mv$, until it reaches its maximum energy at $v\Lambda/2$. In other words, in this regime, the transferred momentum $\delta k$ is mainly absorbed by the particle. The particle can absorb at most momentum $(1-\nu) \Lambda$, and this saturation occurs at the injected momentum $Q_s\!=\!Mv + (1-\nu)\Lambda/2$, fulfilling $E^0(Q_s) = E^{(1-\nu)\Lambda}(Q_s)$; this results in the splitting shown in panels Fig.~\ref{fig:PHsymmetry}(a) and (b). As one increases the injected momentum beyond $Q_s$, the extra momentum goes into the hole cloud, whose occupancy peak shifts until the lower energy bound $-v\Lambda/2$ is reached at $Q\!=\!Q_\Lambda$. Finally, for $Q>Q_\Lambda$, all the single particle-hole states are off-resonant (as already explained above), and the occupancies are small. 

While an analytical understanding remains elusive at this stage, we speculate that the physical mechanism underlying this behavior (in particular the asymmetry in the particle and hole behaviors) is that the impurity tends to bind to holes when $U>0$ and $M>0$. 

The whole picture is reversed in the dual case [Eq.~\eqref{duality_transf}], where particle and hole exchange roles; now the impurity correlates with the particle. As illustrated in panel Fig.~\ref{fig:PHsymmetry}(d), the particle-hole duality expressed by Eq.~\eqref{eq:dualH} leads to an occupancy-plot that corresponds to the mirrored image (vertically mirrored with respect to $\omega_{\mathrm{bath}}\!=\!0$) of that shown in panel Fig.~\ref{fig:PHsymmetry}(c).

We point out that this particle-hole duality provides a useful tool in the analysis of other fermionic systems at half-filling, e.g.~the Fermi-Hubbard model~\cite{amelio2024insulator}. 

\subsection{Lattice effects}

\begin{figure}[t]
\centering
\includegraphics{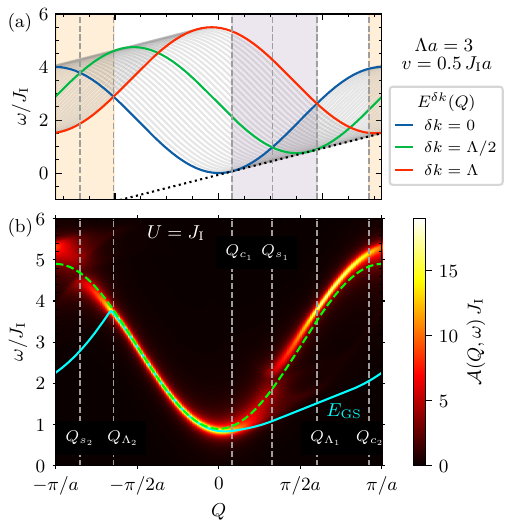} 
\caption{Lattice effects. (a) Non-interacting spectrum of the impurity-bath system as a function of the injected impurity momentum $Q$, parameterized by the transfer momentum $\delta k \in [0,\Lambda]$. (b) Spectral function of an impurity moving on the lattice and interacting with a chiral bath, which has a linear dispersion characterized by its group velocity $v$. Here, we set $\Lambda a\!=\!3$, $v\!=\!0.5J_{\mathrm{I}}a$ and the interaction strength $U\!=\!J_{\mathrm{I}}$ for $N_P\!=\!100$ bath modes at half-filling.
}
\label{fig:trittic_1D_chiral_lattice}
\end{figure}

We finally comment on the consequences of adding another length scale in the problem, associated with the impurity. This can be analyzed, for instance, by setting the impurity on a 1D lattice. In this case, the impurity length scale is given by the lattice constant $a$ and its energy scale by the hopping amplitude $J_{\mathrm{I}}$. The system Hamiltonian would still be described by Eq.~\eqref{eq_chiral1D}, albeit with an impurity dispersion $\varepsilon_{\mathrm{I}}(q)\!=\!-2J_{\mathrm{I}}\cos{(qa)}$, where $q \in (-\pi/a, \pi/a]$ is the Bloch quasi-momentum defined modulo $2\pi/a$.

The addition of a new scale has some consequences. In particular, if the group velocity $v$ of the chiral medium is larger than the maximum group velocity  $\max(|\mathrm{d}\varepsilon_{\mathrm{I}}/\mathrm{d}q|)\!=\! 2J_{\mathrm{I}}a$ allowed for the impurity, then there is no resonant particle-hole state nor critical momentum $Q_c$; in this case, the polaron spectrum will resemble that of the bare impurity. On the other hand, if $v<2J_{\mathrm{I}}a$, then one would obtain two critical momenta $Q_{c_{1,2}}$, for which the impurity group velocity is equal to $v$. These critical momenta are explicitly given by
\begin{equation}
    Q_{c_1} = \frac{1}{a}\sin^{-1}\left( \frac{v}{2J_{\mathrm{I}}a} \right), \quad Q_{c_2} = \frac{\pi}{a} -  Q_{c_1}.
\end{equation}
Hence, in this case, one would find two regions of broadening in the first Brillouin zone, with two corresponding splitting points $Q_{s_{1,2}}$ and cutoff points $Q_{\Lambda_{1,2}}$. Following the same reasoning as in the previous subsection, these critical points are given by the roots of the equation $E^{\delta k}(Q) - \varepsilon_{\mathrm{bath}}(Q)\!=\!0$, where $\delta k\!=\!\nu\Lambda$ for splitting points and $\delta k\!=\!\Lambda$ for the cutoff points.

These predictions are confirmed in Fig.~\ref{fig:trittic_1D_chiral_lattice}. In panel Fig.~\ref{fig:trittic_1D_chiral_lattice}(a), we identify all the critical $Q$-points based on the spectrum of the decoupled impurity-bath system, for $v\!=\!0.5J_{I}a$ and $\Lambda\!=\!3 a^{-1}$. We then show the impurity spectral function in Fig.~\ref{fig:trittic_1D_chiral_lattice}(b), for an interaction strength $U\!=\!J_{\mathrm{I}}$. One clearly recognizes the main features that were already present in the parabolic-dispersion case [see Fig.~\ref{fig:trittic_1D_chiral}(b)], but we also notice the additional broadening features due to the lattice. We note that these additional effects extend to negative $Q$ due to the Bloch periodicity.

\section{Realization of a chiral polaron on the edge of a 2D Chern insulator}
\label{sec:Haldane}

The previous Section built our understanding of the chiral polaron formation, through the analysis of a mobile impurity interacting with a 1D chiral bath. However, in practice, 1D chiral modes can only exist at the boundary of a higher dimensional topological system. In the next Sections, we consider more realistic models for the topological chiral bath, both non-interacting and strongly-interacting.

In this Section, we consider a 2D Chern insulator (CI) of non-interacting fermions, defined on the Haldane model, which hosts a single chiral edge mode within the topological bulk gap. As before, we constrain the impurity to move on the edge of the system, and we numerically calculate its spectral function using MPS time evolution methods and Chevy ansatz. We also use these methods to analyze the dressing of the impurity by the topological chiral edge state.

\begin{figure}[!b]
    \centering
    \includegraphics{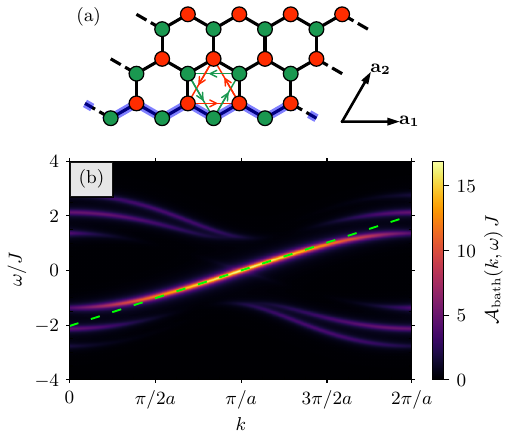}
    \caption{(a) Haldane lattice model, with  $A,B$ sublattice sites denoted by green and red circles, respectively, and nearest neighbour (NN) hoppings  by black lines. The NNN hopping between A lattice sites (red lines) and between B lattice sites (green lines) are shown for a single plaquette; the sign of the corresponding Peierls phase $\varphi$ is indicated by the arrows, illustrating the breaking of time-reversal symmetry. Thick blue lines represent the NN hopping of an impurity along the lower edge of the lattice. {(b)} Single-particle Green's function in Eq.~\eqref{eq:Haldane_spectral_function}, for the non-interacting Haldane model in Eq.~\eqref{eq:Haldane_hamiltonian_real_space}, evaluated along the edge of an infinite strip with a width of 3 unit cells; here we set $J'\!=\! 0.25\,J$ and $\varphi=\pi/2$. The green dashed line is the linear approximation of the chiral mode dispersion, 
    with group velocity $v=4Ja/2\pi$.}
    \label{fig:Haldane_strip}
\end{figure}

\subsection{System Hamiltonian}
We consider a bath of non-interacting spinless fermions in the Haldane model~\cite{haldane1988}, and an impurity confined on the zig-zag edge of the corresponding honeycomb lattice, as schematically shown in Fig.~\ref{fig:Haldane_strip}(a) for an infinite strip.

The total Hamiltonian of the system has the general form of Eq.~\eqref{eq:generalized_system_Hamiltonian}, with the impurity hopping along the edge indicated by the blue chain in Fig.~\ref{fig:Haldane_strip}(a), and with contact interactions between the impurity and the surrounding bath of  fermions. The latter are defined on a honeycomb lattice, as described by the Haldane Hamiltonian~\cite{haldane1988}
\begin{equation}\label{eq:Haldane_hamiltonian_real_space}
\begin{split}
         \hat{H}_{\mathrm{bath}} &\equiv \hat{H}_{\mathrm{Haldane}} = -J \,\sum_{\langle \mathrm{NN} \rangle}  \hat{c}_{xy\sigma}^{\dagger} \hat{c}_{x'y'\sigma'} \\
        & \quad -  J'\,\sum_{\langle\langle \mathrm{NNN}\rangle\rangle} e^{\pm i\varphi}  \,\hat{c}_{xy\sigma}^{\dagger} \hat{c}_{x'y'\sigma}.
\end{split}
\end{equation}
Nearest-neighbor (NN) and next-nearest-neighbor (NNN) hopping amplitudes are denoted as $J$ and $J'$, respectively. Here, $\hat{c}_{xy\sigma}^{\dagger}$ ($\hat{c}_{xy\sigma}$) is the creation (annihilation) operator that creates (destroys) a bath particle at lattice site $\mathbf{r_i} = x\mathbf{a_1} + y\mathbf{a_2}$, and $\sigma=A,B$ denotes the sublattice index.  The sign of the hopping phase $\varphi$  is indicated by the arrows in Fig.~\ref{fig:Haldane_strip}(a). Time-reversal symmetry is explicitly broken and a  topological bulk gap opens in the spectrum of the bath provided that
$\varphi \neq 0,\pi$ and $J' \neq 0$. 
In the following, we will fix the parameters $\phi\!=\!\pi/2$ and $J'\!=\!0.25J$. We will generally consider a cylindrical geometry, or an infinite strip, as represented in Fig.~\ref{fig:Haldane_strip}(a).

Figure~\ref{fig:Haldane_strip}(b) shows the imaginary part of the single-particle Green's function (i.e.~the spectral function) of the bath fermions,
when evaluated on the edge of the system,
\begin{align}\label{eq:Haldane_spectral_function}
    \mathcal{A}_{\mathrm{bath}}(k,\omega) &= -2{\rm Im} \sum_{\sigma\sigma'}\langle 0 | \hat{c}_{k0\sigma} \frac{1}{\omega - \hat{H}_{\mathrm{bath}} + i\eta} \hat{c}_{k0\sigma'}^{\dagger}| 0 \rangle.
\end{align}
Here $k\!=\!0,1,\ldots,N_1 - 1 \times (2\pi/N_1 a)$ is the quasi-momentum along $\mathbf{a_1}$, $N_1$ is the number of unit cells along this direction, $a\!=\!|\mathbf{a_1}|$, $\ket{0}$ is the vacuum state, and
the Fourier transform of the field is defined as
\begin{align}
    \hat{c}_{ky\sigma} &= \frac{1}{\sqrt{N_1}} \sum_{x} e^{-i k \mathbf{r}_i \cdot \mathbf{a}_1} \hat{c}_{xy\sigma}.
    \label{eq:Fourier_c}
\end{align}
This operator removes a bath particle with crystal momentum $k$ along the $x$-direction, at position $y$, on the sublattice  $\sigma$. The spectral function in Eq.~\eqref{eq:Haldane_spectral_function} can be calculated by diagonalizing Eq.~\eqref{eq:Haldane_hamiltonian_real_space} in $k$-space. The presence of a chiral mode of well-localized edge states is clearly visible in Fig.~\ref{fig:Haldane_strip}(b); since the bulk gap is of the order of $4J$, the group velocity $v$ of the edge mode is approximately $4J a/2\pi$ (see the green dashed line). 

We note that the bulk modes have a small oscillator strength, signaling a weak overlap between bulk and edge states. Hence, when the impurity-bath interaction is written in the basis of the bath eigenstates, the coupling to the edge modes is substantially larger than the coupling to bulk modes; see Appendix~\ref{appendix:haldane_chevy_ansatz} for details.

We now calculate the momentum-resolved impurity spectral function $\mathcal{A}(Q,\omega)$ for the coupled impurity-bath system. Here, the impurity is confined to the edge of the Chern insulator:~the bath consists of a Fermi sea of non-interacting fermions, completely filling the lowest Bloch band of the Haldane model. To do so, we use both the Chevy approach on a honeycomb cylinder and the MPS evolution on an infinite  strip. We refer to Appendices \ref{sec:MPS_simulations} and \ref{appendix:haldane_chevy_ansatz}, respectively, for details on these methods.

\begin{figure}[!t]
    \centering
    \includegraphics{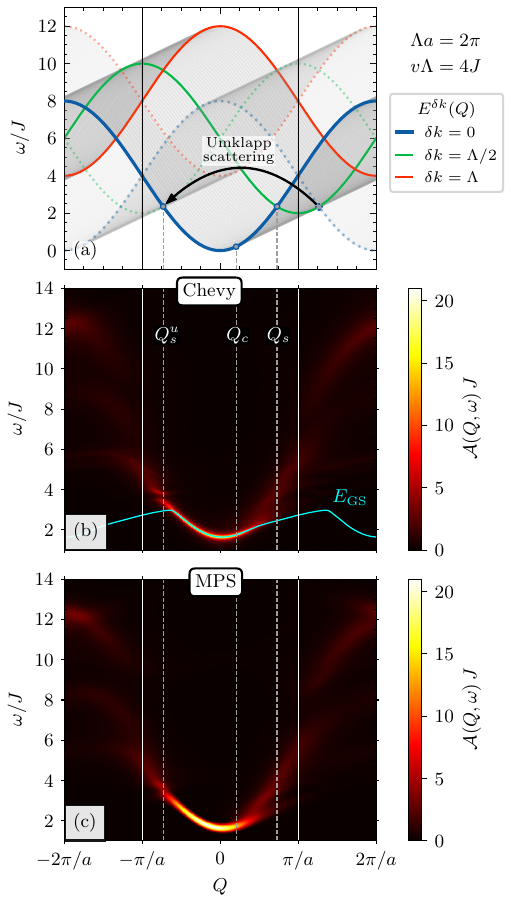}
    \caption{Impurity spectral function in the Haldane model. (a) The non-interacting spectrum 
    $E^{\delta k}(Q)$
    of the system, as a function of the  momentum $Q$ injected by the impurity, and parameterized by the transfer momentum $\delta k$ used to create a single particle-hole pair (grey curves). Momentum conservation is relaxed by the Umklapp processes, which carry momentum $2\pi/a$, giving rise to the  $E^{\delta k}_u(Q)$ curves (light gray and dotted lines). Spectral function of the impurity $\mathcal{A}(Q,\omega)$ calculated using:~(b) the Chevy ansatz, and (c) the MPS time evolution; here, we set $U\!=\!5J$ and the impurity hopping $J_{\mathrm{I}}\!=\!2J$.
    The green dotted curve marks the ground state of the system, which is dark for large momenta.
    The parameters used in the Haldane model are: $J' = 0.25\,J$, and $\varphi = \pi/2$; the Fermi level for the bath particles is set at $0\,J$. For the MPS calculations, the width of the infinite strip is 3 units cells; for the Chevy-ansatz calculations, the cylinder has a height of 3 unit cells and a circumference of 33 unit cells.}
    \label{fig:CI_polaron_chevy_MPS}
\end{figure}

\subsection{Chevy ansatz and MPS results}

The impurity spectral functions obtained from the Chevy ansatz and MPS methods are compared in Fig.~\ref{fig:CI_polaron_chevy_MPS}, for an interaction strength $U\!=\!5J$ and an impurity hopping $J_{\mathrm{I}}\!=\!2J$; here, the fermionic bath is prepared at exactly half-filling. As previously, $Q$ is the momentum injected by the impurity into the system.

As a technical note, we remark that the spectral function in represented in the extended zone scheme. This is motivated by the fact that experiments typically probe the impurity dynamics as a function of momentum rather than quasi-momentum. In the absence of coupling with the bath, the impurity hops on the edge of the lattice, with no bias between $A$ and $B$ sites, such that the natural unit cell for the impurity is half of that of the bath fermions. On the zig-zag edge, the fermionic density on $A$ sites is smaller than on $B$ sites, hence the impurity momentum is only conserved modulo $2\pi/a$; this is the analog of Umklapp processes in solids.

In Fig.~\ref{fig:CI_polaron_chevy_MPS}(a), we show the non-interacting spectrum of the system for different values of the transferred momentum $\delta k$ (gray curves) under the approximation that the Chern-insulator bath consists of a single linearly dispersive mode with group velocity $v\!=\!4Ja/2\pi$ and  cutoff momentum $\Lambda\!=\!2\pi/a$. These non-interacting energy lines are  given by two families of curves 
$E^{\delta k}(Q) \!=\! \varepsilon_{\mathrm{I}}(Q-\delta k) + v\delta k$ 
and 
$E_u^{\delta k}(Q) \!=\!  \varepsilon_{\mathrm{I}}(Q-\delta k+2\pi/a) + v\delta k$, 
where the $E_u$ energies are the ones enabled by Umklapp processes and are depicted in a lighter contrast.  

The chiral features of the impurity spectral function, direct consequence of the interaction with the Chern insulator's chiral edge mode, are clearly visible in Figs.~\ref{fig:CI_polaron_chevy_MPS}(b) and (c). Starting at $Q\!=\!0$, and moving towards the right, one observes that the impurity behaves as a bare particle  until it reaches the critical momentum $Q_c = (2/a)\mathrm{sin}{}^{-1} (v/J_{\mathrm{I}}a)$, at which point the impurity starts to resonantly exchange momentum with the bath through the creation of particle-hole pairs. For $Q>Q_c$, we observe the emergence of a dark ground state (green dots) and the broadening of the spectral lines. Furthermore, a splitting is present at $Q_s$, as the curve $E^{\delta k}$ corresponding to $\delta k \!=\! \Lambda/2$ becomes resonant  with the bare-impurity dispersion (see previous Section); this splitting is hardly visible, due to blurring effects associated with the coupling to the many bulk modes. On the $Q<0$ side, the impurity behaves as a bare particle for small momenta, until it reaches $Q\!=\!Q_s^u$, which corresponds to the resonance with the Umklapp  curve $E^{\delta k}_u$ at $\delta k\!=\!\Lambda/2$.

The identification of these few critical momenta $Q_{c,s}^{(u)}$ in Figs.~\ref{fig:CI_polaron_chevy_MPS} clearly demonstrates how the spectral features of the dressed impurity can be linked to resonances in the non-interacting theory, based on a linearly dispersing chiral mode. In particular, we find that the vertical lines depicted in Fig.~\ref{fig:CI_polaron_chevy_MPS}, which correspond to the critical momenta predicted by neglecting the curvature of the edge-mode dispersion in the non-interacting theory, show good agreement with the impurity spectral function of the fully interacting impurity-bath system. It is also worth noticing that this spectral function was calculated using a large interaction strength $U$, larger than the bulk gap in the bath, such that bulk excitations are potentially relevant; see also the next paragraph. Altogether, this demonstrates that the mechanism underlying the formation of the chiral polaron, as well as its related spectral features, are robust. 

In  Fig.~\ref{fig:CI_chevy_without_bulk} in Appendix~\ref{app:ED}, we provide further details by filtering out the contribution of the bulk  modes, which have the general effect of blurring the spectral features at large momenta; this complementary study leads to a complete identification and classification of all the critical $Q$-points.

\subsection{Polaron formation on the edge vs in the bulk}

\begin{figure}[!t]
    \centering
    \includegraphics{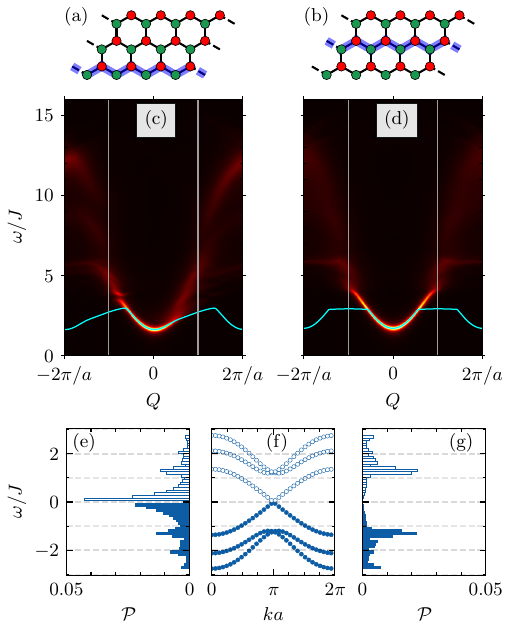}
    \caption{Polaron formation on the edge versus in the bulk of a Chern insulator. The location of the injected impurity is indicated by the blue links in panels (a) and (b), respectively. The corresponding impurity spectral functions $\mathcal{A}(Q, \omega)$, calculated using the Chevy ansatz, are reported in panels (c) and (d). Panels (e) and (g) show the energy-resolved probability distribution for the creation of a bath particle (empty bars) and hole (filled blue bars) in the interacting many-body ground state at $Q\!=\!0$. Panel (f) displays the single-particle band spectrum of the Haldane bath. Here $J_{\mathrm{I}}\!=\!2J$ and $U\!=\!5J$.}
    \label{fig:CI_bulk_edge}
\end{figure}

To further verify that the formation of the chiral polaron is indeed due to the interaction of the impurity with the topological edge mode, we display the impurity spectral function in Fig.~\ref{fig:CI_bulk_edge} for two distinct cases: (i)~when the impurity is constrained to move along the edge [Fig.~\ref{fig:CI_bulk_edge}(c)]; and (ii)~when it is constrained to move along a path within the bulk [Fig.~\ref{fig:CI_bulk_edge}(d)]. The blue lines in panels Fig.~\ref{fig:CI_bulk_edge}(a) and (b) signify the impurity path constraints. Figure~\ref{fig:CI_bulk_edge}(d) clearly illustrates that the impurity spectral function is symmetric whenever the impurity is constrained to a path within the bulk. This confirms that the asymmetry visible in panel Fig.~\ref{fig:CI_bulk_edge}(c) originates from the chiral edge mode. This is further illustrated in panels Fig.~\ref{fig:CI_bulk_edge}(e) and (g), where we plot the probability  of particle (empty bars) and hole (filled bars) excitations at energy $\omega$, for the Chevy wavefunction of the ground state at $Q\!=\!0$. In panel Fig.~\ref{fig:CI_bulk_edge}(f), we plot the corresponding particle (empty circles) and hole (filled circles) bands. Figure~\ref{fig:CI_bulk_edge}(e) shows that, for an impurity  constrained to move on the edge, the polaron dressing cloud mostly consists of low-energy excitations within the chiral branch (i.e.~modes between $\pm J$ approximately); in contrast, when confining the impurity in the bulk, the result in Fig.~\ref{fig:CI_bulk_edge}(g) shows that low-energy excitations are highly suppressed, resulting in a symmetrical spectral function, reminiscent of the non-chiral bath case discussed in Sec.~\ref{sec:1DFermi}.

In Appendix~\ref{app:insulator_vs_metallic}, we also demonstrate that the chirality of the edge mode only clearly manifests itself in the spectral function $\mathcal{A}(Q,\omega)$ if the Fermi level of the bath lies within the topological gap, i.e.~when the bath forms a genuine Chern insulator. Indeed, when the bath is prepared in a metallic phase, the impurity spectral function only displays weak asymmetries and its features are difficult to predict. In this sense, one concludes that polaron spectroscopy is sensitive to the topological nature of the surrounding many-body medium.

\section{Polaron on the edge of a fractional Chern insulator}
\label{sec:fractional}

As a final application of our chiral polaron framework, we study the impurity dynamics on the edge of a fractional Chern insulator (FCI). Similar to the case of the non-interacting Chern insulator (Haldane model) discussed in Sec.~\ref{sec:Haldane}, we expect that the impurity couples to the chiral, linearly dispersing edge mode of the FCI, and that the general features found in the previous Sections extend to this strongly-correlated-bath setting. However, since the bath is indeed an interacting, strongly correlated system, and since we do not have any analytical access to the  ground state of the bath, the Chevy method cannot be used and we entirely resort to  MPS calculations. The numerical results presented in this Section confirm that the chiral polaron picture persists even in the presence of strong quantum correlations in the bath.

\begin{figure}[t]
    \centering
    \includegraphics{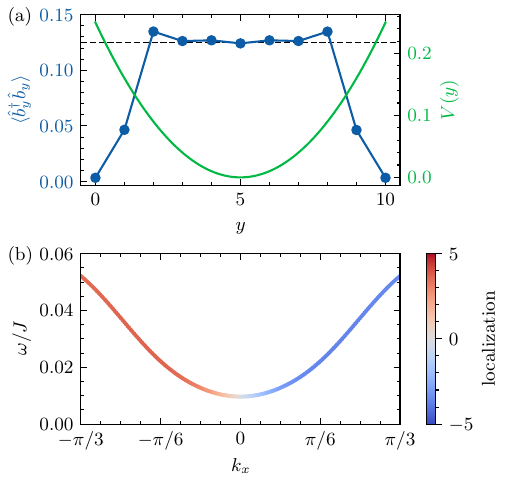}
    \caption{Fractional Chern insulator on a strip. (a) The density profile along the vertical direction of the strip, imprinted by the harmonic confining potential. Note that the sum of the densities on a rung is exactly one, as explicitly encoded in the infinite MPS representation.
   (b) Dispersion of the lowest-energy neutral mode of the FCI, the color indicating the average position along the $y$ direction with respect to the center of the system. The bulk gap is around $0.25J$ for this system configuration (see text).} 
    \label{fig:fci_density}
\end{figure}

We consider the Hofstadter-Bose-Hubbard model on a square lattice, which is known \cite{Sorensen2005, Hafezi2007, Gerster2017, Weerda2023} to host a stable Laughlin-type FCI ground state at filling factor $\nu\!=\!1/2$, for large enough interactions. Here, the filling factor $\nu$ refers to the filling of the lowest-energy (single-particle) Hofstadter band; for convenience, we will consider a bath of hard-core bosons in such a band. This Laughlin-type FCI ground state is known to host a single chiral edge mode on its boundary~\cite{Moore_edge,Binanti2024}.

In our calculations, we consider an infinite strip geometry in order to have a gapless edge mode with a continuous momentum. On the strip, the most convenient gauge choice for the vector potential is $\vec{A}\!=\!(y\varphi,0)$, such that the Hamiltonian is invariant with respect to translations in the horizontal direction:
\begin{multline}
    \hat H_{\text{FCI}} = -J \sum_{xy} \left( e^{i\varphi y} \hat{c}_{x,y}^\dagger \hat{c}_{x+1,y} + \mathrm{h.c.} \right) \\ -J \sum_{xy} \left( \hat{c}_{x,y}^\dagger \hat{c}_{x,y+1}+ \mathrm{h.c.} \right) - \sum_{xy} V(y) \, \hat{n}_{xy} \;.
\end{multline}
with $\hat{n}_{xy}\!=\!\hat{c}_{xy}^\dagger \hat{c}_{xy}$. The $x$ label runs over the infinite direction of the strip, whereas $y\!=\!0,\dots,N_y-1$; in the following we consider a strip of thickness $N_y\!=\!11$. We set the magnetic flux per plaquette at the value $\varphi\!=\!\pi/2$, and impose that there is, on average, one boson per rung in the ground state resulting in a filling  $\nu\!=\!1/2$ of the lowest Hofstadter band.

It is well-known in the literature that a suitable confining potential is crucial for observing a linearly dispersing chiral edge mode in such FQH systems~\cite{Wen1992,Moore_edge,Fern2017, Macaluso2017,Binanti2024}. In particular, we expect that a hard-wall confinement will gap out the edge mode \cite{Fern2017, Macaluso2017,Repellin2020,Binanti2024}. Instead, we use a harmonic potential of the form $V(y) = \mu (y-y_0)^2$, where $y_0$ is the center of the strip. We use the value $\mu/J\!=\!0.01$ for the strength of the potential, which was shown in Ref.~\cite{Binanti2024} to yield a clear signature of the chiral branch.

\begin{figure}[b]
    \centering
    \includegraphics{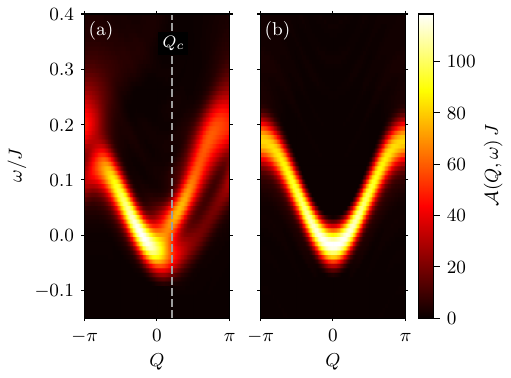}
    \caption{
    Impurity spectral function for an impurity injected in a fractional Chern insulator. The impurity is constrained to move:~(a) on the edge ($y\!=\!2$); (b) in the bulk ($y\!=\!5$). The spectral function is computed by MPS time evolution methods on the infinite strip geometry. Here, we used $J_{\mathrm{I}}\!=\!0.05J$ and $U\!=\!J$. The vertical line depicts the critical momentum $Q_c\!=\!M v$.}
    \label{fig:FCI_bulk_edge_comparison}
\end{figure}

Given this infinite-strip geometry with harmonic potential, we optimize an MPS approximation for the ground state. In Fig.~\ref{fig:fci_density}(a), we observe a clear plateau of the boson density $\braket{\hat{n}_{xy}}$ around the value $n=1/8$ (dashed black line). We have thus realized an FCI ``droplet'' on this infinite strip geometry. 

Next, using a variational MPS approach for the low-lying excited states in 2D correlated systems \cite{VanDamme2021}, we obtain an estimate of the different excitation energies in this system. First of all, by placing the system on an infinite cylinder with a circumference of eight sites, we find that the bulk gap is about $\Delta_{\text{bulk}}\approx0.25J$. Secondly, on the strip geometry with harmonic potential, we find an upper bound for the edge gap $\Delta_{\text{edge}}<0.01J$, and an effective edge velocity $v \approx 0.06J$. 

The localization of the edge mode can be further diagnosed by computing the spectral weight of the low-energy excited state as
\begin{equation}
    s_y(k_x) = \bra{\Phi_{k_x}} \sum_x e^{ik_xx} (\hat{n}_{xy}-\braket{\hat{n}_{xy}}) \ket{\Psi_0},
\end{equation}
where $\ket{\Psi_0}$ is the ground state in the bath, and $\ket{\Phi_{k_x}}$ is the excited state wavefunction with momentum $k_x$, both in their variational MPS representation. From this quantity, one can estimate the average position of the edge mode as 
\begin{equation}
\langle y \rangle\!=\!\sum_y y \left| s_y \right|^2 / \sum_y \left| s_y \right|^2.
\end{equation}
The dispersion and the average position of the edge mode are plotted in Fig.~\ref{fig:fci_density}(b). We find an approximately linear dispersion for a significant portion of the Brillouin zone, where the edge mode is mostly located between sites $y\!=\!2,3$ for positive momenta; for negative momenta, the edge mode is localized at the other side of the strip. Around momentum zero, the dispersion bends off and the edge localization disappears. We attribute this behavior around the origin either to a coupling of the edge modes across the strip, to the finite trapping frequency, or to artifacts in the variational MPS calculations. We leave a more quantitative analysis of the FCI edge spectrum on an infinite strip geometry for future work.

Finally, we immerse a mobile impurity into the FCI droplet, which we restrict to move only horizontally on one row of the strip with hopping strength $J_I$. The total Hamiltonian of the system is given by
\begin{multline}
    \hat H = \hat H_{\text{FCI}} - J_I \sum_{x}  \left( \hat{d}_{x,y}^\dagger \hat{d}_{x+1,y} + \mathrm{h.c.} \right) \\ + U \sum_{x} \left( \hat{c}_{x,y}^\dagger \hat{c}_{x,y} \right) \left(\hat{d}_{x,y}^\dagger \hat{d}_{x,y} \right)
\end{multline}
where $y$ denotes the location of the impurity along the $y$-axis on the strip. We take the parameters $J_I\!=\!0.05 J$ and $U\!=\!J$, which are small enough to expect that the impurity mainly couples to the edge mode, provided that $y$ is chosen to be on the edge of the system. We again use time-dependent MPS methods to compute the polaron spectral function, and the results are shown in Fig.~\ref{fig:FCI_bulk_edge_comparison} for two choices of the impurity location.

In line with the results of the previous Sections, one can retrieve the key characteristics of the chiral polaron in Fig.~\ref{fig:FCI_bulk_edge_comparison}(a), which corresponds to the case of an impurity injected on the edge of the FCI strip. First, we observe that the spectral weight becomes asymmetric due to the coupling to the chiral edge mode. Second, we observe a splitting of the spectral function around the critical momentum $Q_c\!=\!M v$, at which the group velocity of the impurity equals the velocity of the FCI edge mode. In contrast, when the impurity is constrained to move within the bulk of the system, one does not find such a chiral response and the spectral function is symmetric; see Fig.~\ref{fig:FCI_bulk_edge_comparison}(b). This demonstrates that the chiral polaron characteristics observed in Fig.~\ref{fig:FCI_bulk_edge_comparison}(a) indeed result from the coupling of the impurity to the chiral mode located on the edge of the FCI droplet. 

This study illustrates how polaron spectroscopy can be used to probe and characterize the edge-mode structure of strongly-correlated topological matter.

\section{Conclusions and outlook}
\label{sec:conclusions}

This work introduced the chiral polaron, a mobile impurity dressed by a bath of chiral fermions. We hereby  summarize our main results and list a series of open questions and perspectives.\\

First, we have provided a complete calculation of the spectral properties of 1D Fermi polarons on a non-chiral chain, benchmarking the accuracy of the Chevy-ansatz and MPS approaches against exact diagonalization. 
Then, we have introduced an effective 1D model for the chiral polaron, revealing its unique spectral properties inherited from the chirality of the bath.
In particular, the impurity propagation was shown to be strongly modified whenever its group velocity becomes comparable to the one of the chiral modes.
The ground state, corresponding to the impurity being dragged by the chiral bath, then becomes dark.
The role of particle and hole processes was analyzed, and a duality relation was established.
Building on this strong theoretical understanding, we considered a realistic 2D Chern-insulator setting 
provided by the Haldane model, which displays a chiral edge mode propagating along the boundary of the system. 
In this setting, the impurity spectrum inherits the chirality of the edge mode, in spite of  contributions stemming from the bulk modes, and provided that the impurity is confined on the edge of the system. Importantly, these signatures of the chiral polaron disappear when the impurity is injected in the bulk of the system, or when the bath is prepared in a metallic phase. In this sense, the chiral polaron can be used as a practical probe of the topological chiral edge modes characterizing topological insulating states of matter. Finally, we used MPS methods to reveal the formation of a chiral polaron 
on the edge of a fractional Chern insulator, a strongly-correlated topological state of matter.
We remark that the adaptation of advanced tensor network methods to the study of polaron formation constitutes an important outcome of this work.

Our results are relevant to cold-atom experiments, where the momentum-resolved polaron spectral function can be measured through Raman injection spectroscopy~\cite{Ness2020,Diessel_2022}, and where important efforts are devoted to the realization of topological insulating states of matter~\cite{cooper2019}, including fractional Chern insulator states~\cite{leonard2023}.
Moreover, the chiral dressing of excitons or electrons could also be relevant in recent solid-state realizations of (fractional) Chern insulators~\cite{cai2023, zeng2023, park2023, xu2023,liu2024, zhao2024}, even though the momentum-resolved spectral function might not be easily accessible in these settings.

On the theoretical side, this work leaves a number of outstanding questions and possible extensions.
These concern the detailed mechanism underlying chiral polaron formation in fractional Chern insulators; the role of interactions among the constituents of the bath; and the case of multiple chiral edge modes in non-Abelian fractional Chern insulators.
Other intriguing directions of research include the addition of magnetic impurities in spin Hall insulators; the possibility of generating effective polaron-polaron interactions mediated by the chiral bath~\cite{paredes2024perspectiveinteractionsmediatedatoms}; and the emergence of novel many-body phases at the edges or surfaces of topological insulators.

\section*{Acknowledgements}

We are grateful to Leonardo Mazza, Alberto Nardin, Botao Wang, Nader Mostaan, C\'{e}cile  Repellin and Fabian Grusdt for stimulating discussions. Work in Brussels is supported by the ERC Grant LATIS, the EOS project CHEQS and the Fonds de la Recherche Scientifique de Belgique (F.R.S.-FNRS). Computational resources have been provided by the Consortium des Équipements de Calcul Intensif (CÉCI), funded by the Fonds de la Recherche Scientifique de Belgique (F.R.S.-FNRS) under Grant No. 2.5020.11 and by the Walloon Region.
O. K. D acknowledges financial support from the NSF through a grant for ITAMP (Award No: 2116679) at Harvard University. GMB acknowledges the Danish National Research Foundation through the Center of Excellence CCQ (Grant no. DNRF156).

\appendix

\section{MPS simulations}
\label{sec:MPS_simulations}

\newcommand{\diagram}[1]{\;\vcenter{\hbox{\includegraphics[scale=0.5,page=#1]{./Figures/mps_diagrams.pdf}}}\;}

In this Appendix, we give the technical details for our numerical MPS simulations.

\subsection{Setups and initial states}

\begin{figure}
\centering
    \includegraphics[scale=0.4,page=3]{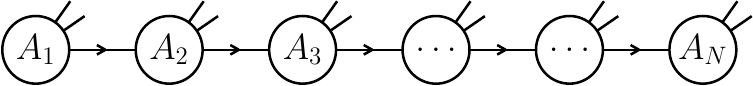}\\\hspace{2cm}\\
    \includegraphics[scale=0.4,page=2]{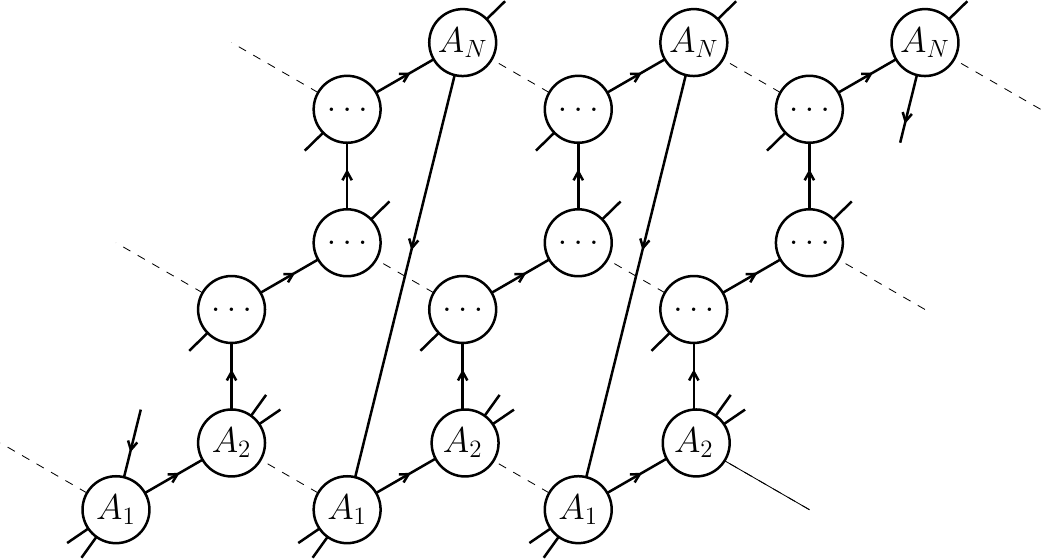}\\\hspace{2cm}\\
    \includegraphics[scale=0.4,page=4]{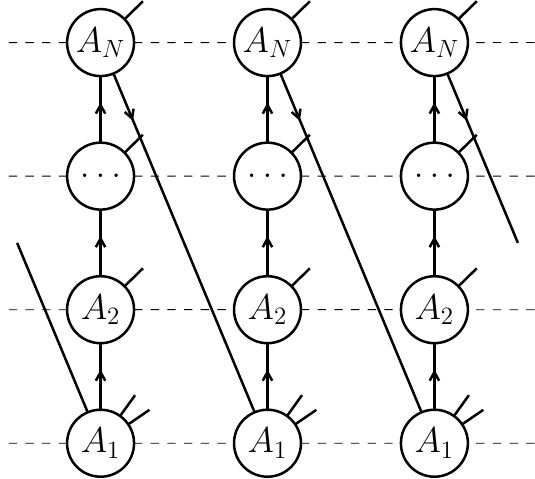}
    \caption{The different geometries considered in this work: a 1D chain, the hexagon lattice on an infinite strip, and the square lattice on an infinite strip. Dotted lines correspond to physical hopping links, while the solid bond with an arrow indicates the 1D MPS representation, winding along the 2D lattice; notice that the two external (physical) legs on the first line denote the presence of the impurity degrees of freedom. The last rows correspond to an infinite MPS.}
    \label{fig:geometries}
\end{figure}

In this work we have considered models on different lattices and with different boundary conditions: a purely 1D chain, a hexagonal lattice on an infinite 2D strip and a square lattice on an infinite cylinder and strip. In the latter cases, we map the 2D geometry to an effective 1D one by winding around the smallest dimension. As a result, we end up with an infinite 1D model with an $N$-site unit cell, but the mapping to an effective 1D geometry induces long-range terms in the Hamiltonian. Besides the geometry, we specify the local Hilbert space. On each site, we have a spinless fermionic or bosonic degree of freedom corresponding to the bath. In addition, we enlarge the local space by an impurity degree of freedom -- no need to specify the statistics for the impurity, because there is always at most one particle in the system. For the 2D geometries, the impurity is restricted to the sites on a 1D slice of the system (typically, the edge). The different geometries and local spaces are represented graphically in Fig.~\ref{fig:geometries}.

For these effective 1D models, we can now propose MPS ans\"atze for approximating the ground state. In the finite cases, we use finite MPS with  a different tensor on each site, whereas the infinite strip requires the use of a translationally invariant MPS, repeating itself within each $N$-site unit cell. We use the DMRG algorithm \cite{White1992, Schollwoeck2011} to variationally optimize the tensors in the finite MPS, whereas we use the VUMPS algorithm \cite{ZaunerStauber2018} to optimize the tensors in the unit cell of the infinite MPS. In all cases we exploit the internal symmetries of the model, namely: 
\begin{itemize}
    \item $\U(1)_{\rm bath}$: charge conservation of the fermionic bath
    \item $\U(1)_I$: charge conservation of the impurity
\end{itemize}
The $\U(1)_{\rm bath}$ symmetry is used to impose the filling of the bath particles. For the finite case, we impose the total number of bath particles by specifying the total charge of all the MPS tensors. In the infinite case, we impose the number of bath particles per unit cell. Initially, there are no impurities in the system, so all MPS tensors only have non-zero entries in trivial sectors of the $\U(1)_I$ symmetry.

\subsection{Impurity spectral function through time evolution with MPS}

In this work we use time dependent MPS methods. First we optimize a ground state MPS for the bath $\ket{\Psi_0}$ with ground state energy $E^0_{\mathrm{bath}}$, apply a local impurity creation operator and perform time evolution of the interacting Hamiltonian, to obtained  the evolved perturbed state
\begin{equation} \label{eq:app_psi_t}
    \ket{\Psi_j(t)} = e^{-i(\hat{H}-E^0_{\mathrm{bath}})t} \hat{d}_j^\dagger \ket{\Psi_0}.
\end{equation}
After each time step, we can compute the correlator straightforwardly
\begin{equation}
    C(j,j',t) = \bra{\Psi_0} \hat{d}_{j'} \ket{\Psi_j(t)}.
\end{equation}
as a function of $j'$. In order to resolve the spectral function, we can first transform to momentum space
\begin{equation}
    \mathcal{A}_j(Q,t) = \sum_{j'} e^{iQ(j-j')} C(j,j',t).
\end{equation}
One can numerically evolve up to  time $T$, which is in practice limited by the bond dimension reachable with the available computational resources, since entanglement increases in time.
Therefore, the Fourier transformation to frequency space requires us to introduce a window function
\begin{equation}
    \mathcal{A}_j(Q,\omega) = \int_{-T}^T \mathcal{A}_j(Q,t) e^{i\omega t} \exp\left( -\frac{t^2}{\alpha T^2} \right)
\end{equation}
where the parameter $\alpha$ can be tuned to give an optimal balance between the final frequency resolution and spurious oscillations.

\subsection{Time evolution in infinite systems}

The idea is that we start from the ground state of the infinite system, and approximate the time-evolved state [Eq.~\ref{eq:app_psi_t}] as a window of time-dependent MPS tensors around site $j$, embedded in the infinite MPS to the left and to the right of the window:
\begin{equation}
    \ket{\Psi_j(t)} \approx \ket{\Psi_{\text{wMPS}}(\{A^l_i\},\{A^r_i\},\{X_i(t)\} ) }
\end{equation}
with
\begin{multline}
    \ket{\Psi_{\text{wMPS}} (\{A^l_i\},\{A^r_i\},\{X_i(t)\} ) }   = \\  \diagram{6}
\end{multline}
Only the tensors $\{X_1,\dots,X_M\}$ are taken to be time-dependent. In each time step, given a set of window tensors $X_i(t)$ we can determine the tensors $X_i(t+dt)$ variationally by optimizing the squared fidelity 
\begin{multline}
    \Big| \hat{U}(dt) \ket{\Psi_{\text{wMPS}} (\{A^l_i\},\{A^r_i\},\{X_i(t)\} ) } - \\ \ket{\Psi_{\text{wMPS}} (\{A^l_i\},\{A^r_i\},\{X_i(t+dt)\} ) } \Big|^2.
\end{multline}
Here, $\hat{U}(dt)=e^{-i(\hat{H}-E^0_{\mathrm{bath}})dt}$ is approximated by a matrix product operator (MPO) \cite{Zaletel2015, VanDamme2023}. Since the window tensors are a string of MPS tensors, a simple sweeping algorithm works very well for optimizing this fidelity in each time step.

We can dynamically grow the size of the time-dependent window in order to capture the spreading of the perturbation as time evolves. In the current context of the impurity problem, we can simply monitor the probability that the impurity has reached the edge of the window, and increase the window size when this probability has reached a certain threshold value.

A crucial advantage of this setup is that we can exploit translation invariance. Indeed, the site $j$ around which the window-MPS is centered is arbitrary, implying that we can just translate the full window-MPS to obtain the time-evolved state centered around site $j'$,
\begin{equation}
    \ket{\Psi_{j'}(t)} = \hat{T}^{j'-j} \ket{\Psi_j(t)}.
\end{equation}
Using this property, we can split the time evolution operator in the time-dependent correlation function:
\begin{equation}
    C(j,j',t) = \braket{\Psi_{j'}(-t/2) | \Psi_j(t/2)}  .
\end{equation}
In practice, we can run two parallel simulations for the forward and backward time evolutions, and compute the shifted overlaps to obtain the correlator. This trick allows us to reach twice the total simulation time with the same resources.

\subsection{Impurity dynamics with long-range hopping}

The standard time-evolution MPS algorithms struggle to make the impurity hop, because most bond expansion methods rely on two-site updates. 

To make the problem transparent, let us explicitly keep track of the $\U(1)_I$ charge on the different legs in the time evolved MPS. In the ground state there are no impurity particles, so on the virtual level of the MPS tensors there are only contributions with a zero $\U(1)_I$ charge. When we insert the impurity particle at site $j$, we therefore obtain the initial MPS
\begin{equation}
    \diagram{1}.
\end{equation}
We have explicitly denoted the local charge sectors $(0,1)$ on the physical legs where the impurity is allowed to hop, whereas on the other sites we only have the $0$ sector.  After a certain time, the impurity particle should be able to move to other sites $j'$ in the lattice. This means that the time evolved MPS will contain contributions of the form
\begin{equation}
    \diagram{2}
\end{equation}
We need to transfer the non-trivial charge through the virtual level of the MPS from site $j$ (where the non-trivial charge is encoded in the external leg) to site $j'$ (where the physical particle resides). As time progresses, the particle travels further through the system, so that the MPS tensors need to be extended progressively with non-trivial $\U(1)_I$ sectors. A two-site scheme, however, will never be able to generate these non-trivial sectors. If one would try to generate such a non-trivial charge sector on the virtual level between sites $j-1$ and $j$ through a two-site update, one would perform the following decomposition
\begin{equation}
    \diagram{4} = \diagram{5}.
\end{equation}
One immediately sees that a non-trivial charge sector cannot be supported through a two-site update, because the physical space on site $j-1$ does not support an impurity particle. Simple bond expansion schemes based on two-site updates will therefore not work for simulating the impurity dynamics in this context.

However, since we know exactly which sectors we need to add to the virtual level of the MPS, there is an easy fix for this problem: instead of using a two-site scheme, we just initialize the MPS tensors within a certain window around site $j$ with these additional charge sectors, and perform a variational one-site scheme over the extended MPS tensors. Put differently, we just add the right charge sectors in the MPS tensors by hand, and use a variational scheme to optimize the enlarged MPS tensors.

\section{Density of single particle-hole states}\label{app:DOS_chiral}

\begin{figure}
\centering \includegraphics{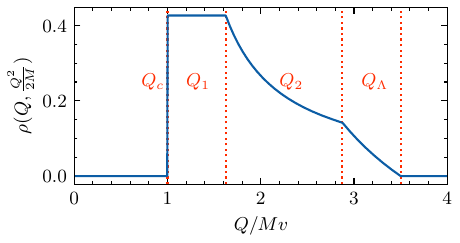}
\caption{
Density of states $\rho(Q,Q^2/2M)$ containing one particle-hole excitation in the $Q$ momentum sector and at kinetic energy resonant with the non-interacting impurity at momentum $Q$. The bath cutoff $\Lambda$ determines the support as well as the kinks in $\rho$.
}
    \label{fig:chevyDOS}
\end{figure}

The density of states in the one particle-hole subspace  at a given bare energy $\omega$ and in the $Q$ momentum sector is given by
\begin{equation}
    \rho(Q,\omega) = \int dk \
    n(k) \
    \delta\left(\omega - \frac{(Q-k)^2}{2M} - vk\right),
\end{equation}
where we introduced the density $n(k)$ of particle-hole pairs with momentum $k$. In other words, $k$ is the momentum which is absorbed by the bath via creation of a particle-hole pair. More precisely,
\begin{equation}
n(k) 
=\frac{1}{Z} \times
\begin{cases}
    k/\Lambda_1 &  {\rm for \ } k \in [0,\Lambda_1] \\
    1  & {\rm for \ } k \in [\Lambda_1,\Lambda_2] \\
    \frac{\Lambda-k}{\Lambda - \Lambda_2} &  {\rm for \ } k \in [\Lambda_2,\Lambda]
\end{cases},
\end{equation}
where $\Lambda_1 = \min (\nu \Lambda, (1-\nu) \Lambda)
,\Lambda_2 = \max (\nu \Lambda, (1-\nu) \Lambda)$,
and $Z=(\Lambda + \Lambda_2 - \Lambda_1)/2$ imposes the normalization $\int dk \ n(k) =1$.

Evaluated at the bare impurity energy, the density of states reads
\begin{equation}
    \rho\left(Q,\frac{Q^2}{2M}\right) = 
    \frac{n(2Q - 2Mv)}{Q-Mv}, 
\end{equation}
and it is shown in Fig.~\ref{fig:chevyDOS} for $\Lambda=4Mv$ and $|\nu-1/2|=1/4$. The density of  states is zero for $Q \leq Q_c = Mv$ or $Q \geq Q_\Lambda = Mv+\Lambda/2$, while it displays two kinks at $Q_1 = Mv+\Lambda_1/2$ and $Q_2 = Mv+\Lambda_2/2$. Notice that  the $Q_s$ splitting point defined in the main text is equal to $Q_2$ for $U>0, \nu<1/2$ or $U<0, \nu>1/2$, and to $Q_1$ in the other cases.

\subsection{Occupation of particle-hole states}
\label{subapp:p-h_probability}

The energy resolved particle-hole occupations
reported in Fig.~\ref{fig:PHsymmetry}(c,d)
are defined 
in terms of the Chevy expansion coefficients
as follows:
\begin{equation}
    \mathcal{P}(Q,\omega_{\rm bath})
    =
    \mathcal{P}_e(Q,\omega_{\rm bath}) +
    \mathcal{P}_h(Q,\omega_{\rm bath}),
\end{equation}
with
\begin{equation}
     \mathcal{P}_e(Q,\omega_{\rm bath})
     =
     \sum_{nkp} |\psi^{Qn}|^2  |\psi^{Qn}_{kp}|^2
     \delta(\varepsilon_{\rm bath}(k) - \omega_{\rm bath}),
\end{equation}
\begin{equation}
     \mathcal{P}_h(Q,\omega_{\rm bath})
     =
     \sum_{nkp} |\psi^{Qn}|^2  |\psi^{Qn}_{kp}|^2
     \delta(\varepsilon_{\rm bath}(p) - \omega_{\rm bath}),
\end{equation}
where 
$n$ labels the many-body eigenstates and $|\psi^{Qn}|^2$ expresses the fact that we are weighting them by their oscillator strength. The idea behind $P$ is to identify which particle and hole states contribute the most to the bright lines in the impurity spectral function.

\section{1D chiral polaron: ED benchmark}
\label{app:ED}

\begin{figure}
    \centering
    \includegraphics{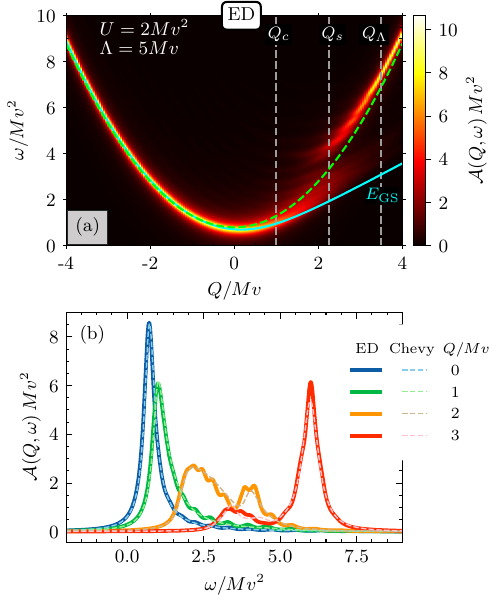}
    \caption{(a) Spectral function of an impurity, with a parabolic dispersion, interacting with a 1D chiral bath calculated using ED for interaction strength $U=Mv^2$ and $\Lambda=4Mv$ at half-filling with the number of bath modes $N_P=20$. Cyan solid curve marks the ground state energy $E_{\mathrm{GS}}$. (b) Comparison of spectral function calculated using ED (solid curves) and Chevy ansatz (dashed curves) for the pivotal $Q$-points indicated by dotted gray lines in panel (a) and $Q=0$.}
    \label{fig:chiral_ED_benchmark}
\end{figure}

In this appendix, we benchmark the calculation of impurity spectral function $A(Q,\omega)$ with a parabolic dispersion in Sec.\ref{sec:1D_chiral} by exactly diagonalizing the Hamiltonian of the system (Eq.\ref{eq_chiral1D}) in the full Hibert space using Krylov subspace methods.
Since the chiral dispersion needs to be represented in a momentum space basis, the interaction term in the Hamiltonian is much denser than the very sparse real space matrices used for the conventional Fermi polaron in Section \ref{sec:1DFermi} which highly restricts the accessible Hilbert space sizes. 
It includes all the higher order processes as opposed to including only one particle-hole pair creation process used in Chevy ansatz.
In Fig.~\ref{fig:chiral_ED_benchmark}(a) we present results for $N=10$ fermions at half filling, meaning $N_P=20$ with the interaction strength $U=Mv^2$ and bath momentum cutoff $\Lambda=4Mv$. In panel (b) we can clearly see a strong agreement between the predictions of the Chevy ansatz and the results obtained from exact diagonalization in the spectral peak positions apart from the minute finite size effects in the ED calculation. Here, we once again see that the influence of higher order particle-hole pair creation processes on the impurity spectral function is very minimal and justifies our use of Chevy ansatz and kinematic arguments based on three-particle elastic scattering to explain the physical origins of the pivotal $Q$-points in the impurity spectrum.

\section{Details of the Chevy ansatz calculation in the Haldane bath case}
\label{appendix:haldane_chevy_ansatz}

The Hamiltonian of the system has the general form of Eq.~\ref{eq:generalized_system_Hamiltonian} where the bath Hamiltonian is given by Eq.~\eqref{eq:Haldane_hamiltonian_real_space} and for an impurity constrained to hop along the edge (blue lines in Fig.~\ref{fig:Haldane_strip}(a)) of the honeycomb lattice, its Hamiltonian reads
\begin{multline}\label{eq:Impurity_Hamitlonian_real_space}
    \hat{H}_{\mathrm{imp}} = - J_{\mathrm{I}} \sum_{x} \bigg[ \left(\hat{d}_{x0A}^{\dagger} + \hat{d}_{x+1,0,A}^{\dagger} \right)\hat{d}_{x0B} + \\
    + \hat{d}_{x0B}^{\dagger}\left(\hat{d}_{x0A} + \hat{d}_{x+1,0,A} \right)  - 2 \, \sum_{\sigma} \hat{d}^{\dagger}_{x0\sigma} \hat{d}^{\dagger}_{x0\sigma} \bigg],
\end{multline}
where $\hat{d}_{xy\sigma}^{\dagger}$ is the impurity creation operator at lattice position $\mathbf{r_i} = x\mathbf{a_1} + y\mathbf{a_2} + \mathbf{\delta}^{\sigma}$, with $x \in \{0,N_1-1\}$ and $y$ constrained to zero, and $J_{\mathrm{I}}$ is the impurity hopping strength. The third  term is added to set the energy minimum of the impurity dispersion at zero.
This is obtained by going to the quasi-momentum representation
\begin{align}
    \hat{H}_{\mathrm{imp}} &= - 2 J_{\mathrm{I}} \sum_{k} \left[ \cos{\left(\frac{k}{2} \right)} - 1 \right]  \bigg[  \hat{d}_{k0A}^{\dagger} \hat{d}_{k0B} + \hat{d}_{k0B}^{\dagger} \hat{d}_{k0A} \bigg].
\end{align}
where
$\hat{d}_{ky\sigma} = \frac{1}{\sqrt{N_1}} \sum_{x} e^{-i k \mathbf{r}_i \cdot \mathbf{a}_1} \hat{d}_{xy\sigma}$
and $q \in (-\pi/a, \pi/a]$ is the quasi-momentum.
This yields two bands of energy
\begin{align}
    \varepsilon_{\mathrm{I}}^{\nu}(q) &= -2J_{\mathrm{I}}\bigg[\nu \cos \left( \frac{q a}{2}\right) - 1 \bigg],
\end{align}
where   $\nu = \pm$ is the band label. Notice that, for a non-interacting impurity, the folding of the cosine impurity dispersion in the halved Brillouin zone is purely formal, having distinguished the equivalent sublattices $A$ and $B$.
However, when the interaction with the bath is turned on, it will be clear that $A$ and $B$ sites are inequivalent, e.g. they have different fermionic density. In the following, it is convenient to keep the indices $\sigma$, instead of using the representation with $\nu$.
The impurity-fermion interaction is contact-like and, keeping into account the confinement of the impurity on $y=0$ sites, it reads
\begin{align}\label{eq:CI_interaction_hamiltonian}
    \hat{H}_{\mathrm{int}} &= U \sum_{x\sigma}  \hat{c}_{x0\sigma}^{\dagger} \hat{c}_{x0\sigma}
    \hat{d}_{x0\sigma}^{\dagger} \hat{d}_{x0\sigma}.
\end{align}

The Chevy approach, which we detail below, is very similar in spirit to the one explained in the main text for the 1D Fermi polaron.
The only nontrivial technical aspect is the implementation of Bloch theorem, which requires dealing with band and quasi-momentum indices.
The Haldane Hamiltonian on a cylinder can be diagonalized in quasi-momentum space to yield
$\hat{H}_{\mathrm{Haldane}} = \sum_{k\alpha} \varepsilon_{\mathrm{bath}}^{\alpha}(k) \, \hat{c}_{k\alpha}^{\dagger} \hat{c}_{k\alpha}$,
where $\alpha = 0,1 \ldots, 2N_2 - 1$ is the band index and $\hat{c}_{k\alpha}^{\dagger} = \sum_{y\sigma} \phi^k_{\alpha, y\sigma} \hat{c}_{ky\sigma}^{\dagger}$  creates a fermion in the eigenstate 
of energy $\varepsilon_{\mathrm{bath}}^{\alpha}(k)$ and wavefunction $\phi^k_{\alpha, y\sigma}$. The ground state of the bath is  the Fermi sea state 
\begin{equation}
    |\Psi_0\rangle \equiv |\mathrm{FS}\rangle = \prod_{\varepsilon_{\mathrm{bath}}^{\alpha}(k) < \epsilon_{\mathrm{F}} } \hat{c}^\dagger_{k\alpha} |0\rangle,
\end{equation}
where  $\epsilon_{\rm F}$ is the Fermi energy. In this basis, the impurity-bath interaction reads
\begin{align}
    \hat{H}_{\mathrm{int}} &= \sum_{\substack{kk'q, \\ \alpha \beta \sigma}}U_{kk'}^{\alpha \beta \sigma} \, \hat{c}_{k\alpha}^{\dagger} \hat{c}_{k'\beta} \, \hat{d}_{q0\sigma}^{\dagger} \hat{d}_{q + k -k',0,\sigma},
\end{align}
where the interaction matrix elements are defined as
\begin{align}
    U_{kk'}^{\alpha\beta \sigma} &= \frac{U}{N_{1}} (\phi_{\alpha, 0\sigma}^{k } )^\star
     \ \phi_{\beta,0\sigma}^{k'}.
\end{align}
In other words, the interaction matrix elements are determined by the localization of the Haldane eigenstates on the edge. One can now project the  Hamiltonian on the truncated Hilbert space of single particle-hole pair excitations in the symmetry sector of injection quasi-momentum $q$, where the states take the Chevy ansatz form
\begin{align}
    |\Psi^q\rangle &= \Bigg(\sum_{\sigma} \psi_{\sigma}^q \,\hat{d}_{q0\sigma}^{\dagger} + \nonumber\\
    &\quad +  \sum_{k k'\sigma}\sum_{\substack{\alpha \notin \mathrm{FS}, \\ \beta \in \mathrm{FS}}} \psi_{k k' \alpha \beta \sigma}^{q}\, \hat{d}_{q+k'-k,0,\sigma}^{\dagger} \hat{c}_{k\alpha}^{\dagger}  \hat{c}_{k'\beta} \Bigg) |\mathrm{FS}\rangle.
\end{align}
The Schr\"odinger equations in this subspace read
\begin{subequations}
\begin{align}
    E\psi_{kk'\alpha\beta\sigma}^{q} &=  \left[\varepsilon_{\mathrm{bath}}^{\alpha}(k) - \varepsilon_{\mathrm{bath}}^{\beta}(k') + E^\sigma_{\rm MF} \right]\psi_{kk'\alpha\beta\sigma}^{q} \nonumber \\
    & \quad - 2 J_{I} \cos \left( \frac{q - k + k'}{2} \right) \bigg[\delta_{\sigma,A} \psi_{kk'B\alpha\beta}^{q} \nonumber \\
    & \quad + \delta_{\sigma B} \psi_{kk'A\alpha\beta}^{q}  \bigg] + \sum_{m\alpha'}\, U_{km}^{\alpha\alpha'\sigma} \psi_{mk'\alpha'\beta\sigma}^{q}\, \nonumber \\
    & \quad - \sum_{m'\beta'} U_{m'k'}^{\beta'\beta\sigma}\psi_{k,m'\alpha\beta'\sigma}^{q} + U_{kk'}^{\alpha\beta\sigma}\, \psi_{\sigma}^{q} \\
    E\psi_{\sigma}^{q} &= E^\sigma_{\mathrm{MF}} \psi_{\sigma}^{q} - 2 J_{I} \cos\left( \frac{q}{2} \right) \Big[ \psi_{B}^{q} \delta_{\sigma, A} + \psi_{A}^{q} \delta_{\sigma, B}\Big]  \nonumber \\
    & + \sum_{kk'\alpha\beta} \quad U_{k'k}^{\beta\alpha\sigma} \psi_{kk'\alpha\beta\sigma}^{q},
\end{align}
\end{subequations}
where crucially the Hartree term $E^\sigma_{\rm MF} = \sum_{m\gamma}\, U_{mm}^{\gamma\gamma\sigma}$ depends on the sublattice $\sigma$, since at the edge of the system the density is not uniform on $A$ and $B$ sites. This term can induce Umklapp scattering, which results in the small gaps visible at $Q=\pm \pi/a$ in the spectra of Fig.~\ref{fig:CI_polaron_chevy_MPS}, as well as in the emergence of the $Q_s^u$ point. 
Note that we have subtracted the energy of Fermi sea $E_{\mathrm{bath}}^0 = \sum_{k,\beta \in \mathrm{FS}} \varepsilon_{\mathrm{bath}}^{\beta}(k)$.

\begin{figure}[b]
    \centering
    \includegraphics{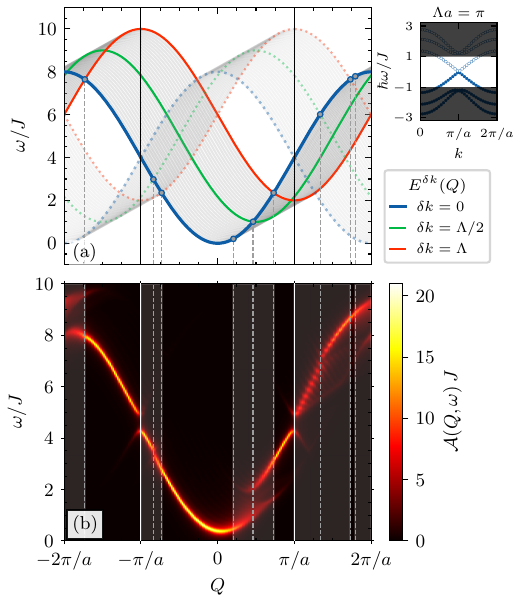}
    \caption{(a) Non-interacting spectrum of the system $E^{\delta k}(Q)$ calculated using the effective chiral model with the linear bath spectrum momentum cutoff $\Lambda=\pi/a$ and group velocity $v=4Ja/2\pi$. The Umklapp scattering enabled spectrum $E^{\delta k}_u(Q)$  is depicted in lighter shade. (b) Impurity spectral function calculated using Chevy ansatz and neglecting all the interaction processes which lead to creation of particles or holes with $|\omega_{\mathrm{bath}}| > J$ (
    upper right inset). Here $J_{\mathrm{imp}}=2J$, $U=5J$ and the Haldane parameters are same as in Sec.~\ref{sec:Haldane}}
    \label{fig:CI_chevy_without_bulk}
\end{figure}

\begin{figure*}[t]
    \centering
    \includegraphics[width=0.9\textwidth]{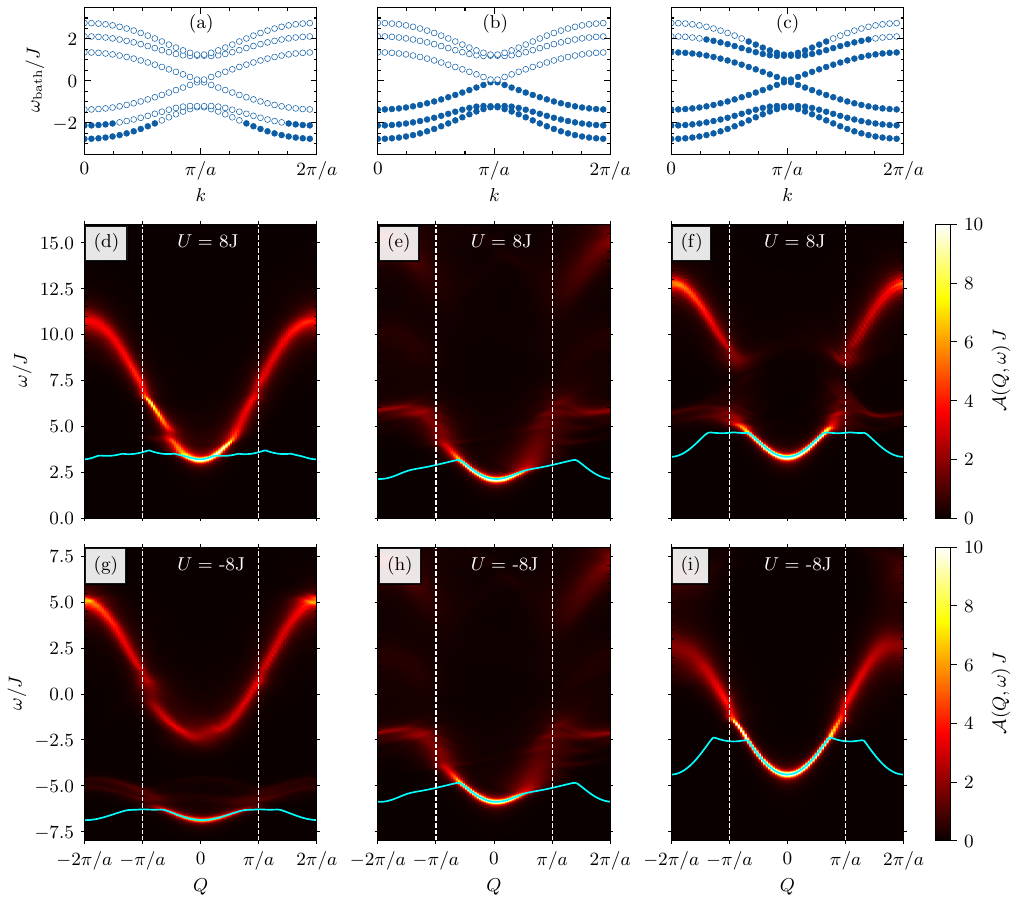}
    \caption{
    Variation in spectral function of impurity wrt Fermi level and sign of interaction strength using Chevy ansatz. Columns correspond to variation in Fermi level of the bath as can be inferred from panel (a)-(c) which depicts the Haldane spectrum with Fermi level at $-2J$ (a), $0J$ (b) and $2J$ (c). $\mathcal{A}(Q,\omega)$ calculated using Chevy ansatz with repulsive (in (d)-(f)) and attractive (in (g)-(i)) on-site interaction between the impurity and the bath particles. Here $J_{\mathrm{imp}}=2J$ and the parameters used in Haldane model are the same as in Sec.~\ref{sec:Haldane}.}
\label{fig:CI_fermi_level_variation}
\end{figure*}

Importantly, we consider the impurity spectral function with momentum defined in the extended-zone scheme.
This is because, in an experiment, the non-interacting impurity will be prepared with momentum $Q \in (2\pi/a, 2\pi/a]$, defined with respect to the 1D chain with equivalent $A$ and $B$ sublattices (as opposed to the quasi-momentum $q$). In other words, in the extended-zone scheme the dispersion of the bare impurity
manifests itself as an unfolded, conventional cosine.
More precisely, the spectral function reads
\begin{equation}
    \mathcal{A}(Q,\omega) =
    -2{\rm Im}
    \langle \Psi_0 | \hat{d}_{Q}
\frac{1}{\omega - \hat{H} + E_{\mathrm{bath}}^0 + i\eta}
    \hat{d}^\dagger_{Q}
    | \Psi_0 \rangle,
\end{equation}
where $\hat{d}_{Q} = \frac{1}{\sqrt{2N_1}} \sum_{x\sigma} e^{-i Q a (x + \frac{1}{2} \delta_{\sigma B})} \hat{d}_{x0\sigma}$.
For $Q=q$ in the first Brillouin zone, one simply has
$\hat{d}_{Q} = \frac{\hat{d}_{q0A} + \hat{d}_{q0B}}{\sqrt{2}}$, while, for $Q=q+2\pi/a$, the sign of the $B$ contribution is flipped, 
$\hat{d}_{Q} = \frac{\hat{d}_{q0A} - \hat{d}_{q0B}}{\sqrt{2}}$.
With this in mind, the spectral function is computed  by numerically diagonalizing the Chevy-Sch\"odinger equations and invoking Eq.~\ref{eq:Aw_En}.

\section{Role of the Haldane bulk modes}

Below we try to better understand the role of the bulk modes first by artificially excluding them from the dressing of the impurity, and then by moving the Fermi level from the topological gap into the Haldane bands, in order to explore metallic states of the bath.

\subsection{Exclusion of the bulk modes}

Since in Fig.~\ref{fig:CI_polaron_chevy_MPS}, the impurity spectral function at large $Q$ values becomes blurred because of interaction with the non-localized bulk modes, it becomes difficult to isolate all the crucial $Q$-points which one can obtain from the effective chiral model. Hence, in Fig.~\ref{fig:CI_chevy_without_bulk} we filter out all the interaction processes in the Chevy ansatz calculation which lead to creation of an electron or hole with energy modulus greater than $J$ (black shaded area in the bath spectrum plot on top right of panel (a)) for the same value of $J_{\mathrm{imp}}=2J$ and $U=5J$ at half-filling, though in this case the bath cutoff momentum is reduced to $\pi/a$. 
The resulting non-interacting spectrum $E^{\delta k}(Q)$ is shown in panel (a) (gray curves) and the Umklapp scattering enabled spectrum $E^{\delta k}_u(Q)=E^{\delta k}(Q+2\pi/a)$ in lighter shade. All the crucial $Q$-points are marked by the vertical gray dashed lines and are calculated by the intersection of green and red curves (solid and dotted) with the blue solid curve (conventional impurity dispersion).

\subsection{Investigation of metallic phases}
\label{app:insulator_vs_metallic}

While in the main text we set the Fermi level at zero energy and considered the half-filled Haldane model, it is interesting to consider the scenario where the bulk bands are doped.
In this case, one does not have a topological insulator, but a metal with broken time-reversal symmetry.
In Fig.~\ref{fig:CI_fermi_level_variation} this setting is studied for three fillings, corresponding to hole-doping, half-filling and particle-doping, as sketched in panels (a-c).
The Fermi energies are set to $-2J, 0, 2J$, respectively.
The corresponding spectral functions are computed in panels (d-f) for repulsive $U$ and in (g-i) for attractive $U$. While some asymmetry in $Q$ is present also in the metallic phases, in the topological insulator phase one has the clearest asymmetry, with features clearly related to the ones studied in Sec.~\ref{sec:1D_chiral}.
Another interesting remark is that the particle-hole symmetry of the Haldane model at half-filling results in identical spectra at opposite $U$ values, but for a mean-field shift of $U$. This is a consequence of the particle-hole duality analyzed in Sec.~\ref{ssec:1D_chiral_PH_symmetry} and is visible in panels (e) and (h).

\bibliography{refs}

\end{document}